\def\cm2{cm$^{-2}$}

\def\c2{C~{\sc ii}}
\def\c4{C~{\sc iv}}
\def\fe2{Fe~{\sc ii}}
\def\fe3{Fe~{\sc iii}}
\def\mg1{Mg~{\sc i}}
\def\mg2{Mg~{\sc ii}}
\def\si2{Si~{\sc ii}}
\def\si4{Si~{\sc iv}}
\def\al2{Al~{\sc ii}}
\def\al3{Al~{\sc iii}}
\def\o1{O~{\sc i}}
\def\n1{N~{\sc i}}
\def\h1{H~{\sc i}}

\def\approxlt{\mathrel{\spose{\lower 3pt\hbox{$\sim$}}
        \raise 2.0pt\hbox{$<$}}}
\def\approxgt{\mathrel{\spose{\lower 3pt\hbox{$\sim$}}
        \raise 2.0pt\hbox{$>$}}}

\documentclass[article]{gwp80}
\usepackage{graphicx}
\usepackage{amssymb}
\tabletypesize{\normalsize}  

\def\plottwo#1#2{\centering \leavevmode
\includegraphics[width=.45\columnwidth]{#1} \hfil
\includegraphics[width=.45\columnwidth]{#2}}

\shortauthors{Nemec, Smolec, Benk\H o {\it et al.}}
\shorttitle{Non-Blazhko RR~Lyrae stars in the {\it Kepler} field}
\begin{document}
\large
\pagenumbering{arabic}
\setcounter{page}{84}

\title{Non-Blazhko RR~Lyrae Stars Observed with the \\ \\ KEPLER Space Telescope}

%
%
\author{{\noindent J.M.~Nemec{$^{\rm 1,2}$}, 
R.~Smolec{$^{\rm 3}$},   
J.~M.~Benk\H{o}{$^{\rm 4}$},
P. Moskalik$^{\rm 5}$,
K. Kolenberg$^{\rm 3,6}$, 
R. Szab\'o$^{\rm 4}$,   
D. W. Kurtz$^{\rm 7}$, 
S. Bryson$^{\rm 8}$,
E. Guggenberger$^{\rm 3}$,
M. Chadid$^{\rm 9}$,
Y.-B. Jeon$^{\rm 10}$,
A. Kunder$^{\rm 12}$,
A. C. Layden$^{\rm 13}$,
K. Kinemuchi$^{\rm 8}$,
L. L. Kiss$^{\rm 4}$,
E. Poretti$^{\rm 14}$,
J. Christensen-Dalsgaard$^{\rm 11}$, 
H. Kjeldsen$^{\rm 11}$,
D. Caldwell$^{\rm 15}$,
V. Ripepi$^{\rm 16}$,
A. Derekas$^{\rm 4}$,
J. Nuspl$^{\rm 4}$,
F. Mullally$^{\rm 15}$,
S.~E.~Thompson$^{\rm 15}$, 
W. J. Borucki$^{\rm 8}$
\\
\\
{\it (1) Department of Physics \& Astronomy, Camosun College, Victoria, BC,  Canada;  \\
(2) International Statistics \& Research Corporation, PO Box 39, Brentwood Bay, BC,  Canada; \\
(3) Institut f\"ur Astronomie, University of Vienna, T\"urkenschanzstrasse 17, A-1180 Vienna, Austria; \\
(4) Konkoly Obs.Hung.Acad.Sciences, Konkoly Thege Mikl\'os \'ut 15-17, H-1121 Budapest, Hungary;\\
(5) Copernicus Astronomical Center, ul. Bartycka 18, 00-716 Warsaw, Poland;\\
(6) Harvard College Observatory, 60 Garden Street, Cambridge, MA 02138, USA;\\
(7) Jeremiah Horrocks Institute of Astrophysics, Univ. Central Lancashire, Preston PR1 2HE, UK;\\
(8) NASA Ames Research Center, MS 244-30, Moffett Field, CA 94035, USA;\\
(9) Obs. C\^ote d'Azur, Univ.Nice Sophia-Antipolis, UMR 6525, Parc Valrose, France;\\
(10) Korea Astronomy and Space Science Institute, Daejeon, 305-348, Korea;\\
(11) Department of Physics and Astronomy, Aarhus University, DK-8000 Aarhus C, Denmark;\\
(12) Cerro Tololo Inter-American Observatory, Cassila 603, La Serena, Chile;\\
(13) Physics \& Astronomy Dept., Bowling Green State University, Bowling Green, Ohio OH 43403;\\ 
(14) Osservatorio Astronomico di Brera, Via E. Bianchi 46, 23807 Merate, Italy;\\
(15) SETI Institute/NASA Ames Research Center, MS 244-30, Moffett Field, CA 94025, USA;\\ 
(16) INAF-Osservatorio Astronomico di Capodimonte, Via Moiariello 16, I-80131, Napoli, Italy.
}
}
}
%
%


\begin{abstract}
This paper summarizes the main results of our recent study of the non-Blazhko RR~Lyrae stars observed with the {\it Kepler} space telescope.  These stars offer the
opportunity for studying the stability of the pulsations of RR~Lyrae stars and for providing a reference against 
which the Blazhko RR~Lyrae stars can be compared.   
Of particular interest is the stationarity of the low-dispersion ($\sigma < 1$mmag) light curves 
constructed from $\sim$18000 long-cadence (30-min) and (for FN~Lyr and AW~Dra) the $\sim$150000 
short-cadence (1-min) photometric data points.   Fourier-based [Fe/H] values and other physical characteristics 
are also derived.   When the observed periods are compared with
periods computed with the Warsaw non-linear convective pulsation code better agreement
is achieved assuming {\it pulsational} $L$ and $\cal M$ values rather than the (higher) 
{\it evolutionary} $L$ and $\cal M$ values.

\end{abstract}

\section{Introduction}

In the 105 deg$^2$ field of the {\it Kepler} space telescope there are $\sim$40 RR~Lyr stars brighter than 
17th magnitude being observed at both long and short cadence ({\it i.e.}, integrated exposure times of $\sim$30-min and 1-min, 
respectively).  With the notable exception of RR~Lyrae itself (Kolenberg {\it et al.} 2010, 2011) little was known about the 
stars prior to the launch of telescope.   
Several of the RR~Lyrae stars are c-type stars pulsating in the first overtone mode (Moskalik {\it et al.} 2011, in preparation).
The remainder are ab-type stars pulsating in the fundamental mode, approximately half of which exhibit the Blazhko effect (see Benk\H o {\it et al.} 2010) 
and show phase 
and amplitude variations of the light curves occurring on time scales of $\sim$10 to hundreds of days.
The other half of the RRab stars appear to oscillate in a remarkably constant manner and are the subject
of our recent paper ``Fourier analysis of non-Blazhko ab-type RR Lyrae stars observed with the 
{\it Kepler} space telescope'' (Nemec {\it et al.} 2011), the results of which are summarized here. 
Elsewhere in these conference proceedings Kinemuchi (2011) reviews the general properties of the {\it Kepler} RR~Lyrae star
observations, and Kolenberg (2011) discusses the Blazhko RR~Lyrae stars.

\begin{deluxetable*}{lcclclr}
\tabletypesize{\small}
\tablecaption{Non-Blazhko ab-type RR~Lyrae stars in the {\it Kepler} field: \\ Pulsation Periods, Times of Maximum Light, and
Analyzed Data}
\tablewidth{0pt}
\tablehead{ \\ \colhead{Star} & \colhead{KIC}   & \colhead{ {$\langle Kp \rangle $}   } &
       \colhead{  Period  } & \colhead{ $t_0$ } & \colhead{Kepler } & \colhead{No.}  \\
 \colhead{Name} &  \colhead{Number} &   \colhead{[mag]} & \colhead{[day]} & \colhead{[BJD]} & \colhead{data } & \colhead{Pts.} \\
 }
\startdata
NR~Lyr          & 3733346& 12.683 & 0.6820264(2)  & 54964.7381 & LC:Q1-Q5 & 18333      \\ 
V715~Cyg        & 3866709& 16.265 & 0.47070494(4) & 54964.6037 & LC:Q1-Q5 & 18374      \\                      
V782~Cyg        & 5299596& 15.392 & 0.5236377(1)  & 54964.5059 & LC:Q1-Q5 & 18381      \\
V784~Cyg        & 6070714& 15.370 & 0.5340941(1)  & 54964.8067 & LC:Q1-Q5 & 18364      \\
KIC~6100702     & 6100702& 13.458 & 0.4881457(2)  & 54953.8399 & LC:Q0-Q4 & 14404     \\
NQ~Lyr          & 6763132& 13.075 & 0.5877887(1)  & 54954.0702 & LC:Q0-Q5 & 18759      \\
FN~Lyr          & 6936115& 12.876 & 0.52739845(1) & 54953.2690 & SC:Q0+Q5 &149925     \\
                &   -    &   -    & 0.527398471(4)& 54953.2690 & LC:Q1-Q5 & 18338     \\
KIC~7021124     & 7021124& 13.550 & 0.6224926(7)  & 54965.6471 & LC:Q1    &  1595         \\
KIC~7030715     & 7030715& 13.452 & 0.6836137(2)  & 54953.8434 & LC:Q0-Q5 & 18802      \\
V349~Lyr        & 7176080& 17.433 & 0.5070740(2)  & 54964.9555 & LC:Q1-Q5 & 18314      \\
V368~Lyr        & 7742534& 16.002 & 0.4564851(1)  & 54964.7828 & LC:Q1-Q5 & 18273      \\
V1510~Cyg       & 7988343& 14.494 & 0.5811436(1)  & 54964.6695 & LC:Q1-Q5 & 18394      \\
V346~Lyr        & 8344381& 16.421 & 0.5768281(1)  & 54964.9211 & LC:Q1-Q5 & 18362      \\
V350~Lyr        & 9508655& 15.696 & 0.5942369(1)  & 54964.7795 & LC:Q1-Q5 & 18326      \\
V894~Cyg        & 9591503& 13.293 & 0.5713866(2)  & 54953.5627 & LC:Q1-Q5 & 18362      \\
V2470~Cyg       & 9947026& 13.300 & 0.5485894(1)  & 54953.7808 & LC:Q0-Q5 & 18794      \\
V1107~Cyg       &10136240& 15.648 & 0.5657781(1)  & 54964.7532 & LC:Q1-Q5 & 18373      \\
V838~Cyg        &10789273& 13.770 & 0.4802799(1)  & 54964.5731 & LC:Q1-Q5 & 18241      \\
AW~Dra          &11802860& 13.053 & 0.6872158(6)  & 54954.2160 & SC:Q0    & 14240         \\
                &    -   &   -    & 0.687217(1)   & 54954.2160 & LC:Q1    &  1614          \\     
                &    -   &   -    & 0.6872158(2)  & 54954.2160 & LC:Q5    &  4474          \\
                &    -   &   -    & 0.68721632(3) & 54954.2160 & SC:Q5    &135380        \\
\enddata  
\end{deluxetable*}

After deriving pulsation periods, times of maximum light, and total amplitudes, and characterising the light curves using Fourier
decomposition methods, attention is turned to [Fe/H] values and how the stars compare with other RR~Lyrae
stars in our Galaxy and in the Large Magellanic Cloud.  After establishing that there are systematic offsets between 
the $V$ and $Kp$ Fourier parameters, the $Kp$ versions of the correlations
established by  Kov\'acs, Jurcsik, Walker, Sandage and others are used to derive
physical characteristics for the stars.  These include the dereddened colour $(B-V)_0$, mean effective temperature, $T_{\rm eff}$, 
surface gravity, log $g$, and pulsational and evolutionary luminosities ($L$) and masses ($\cal M$).
Finally, the results are compared with new theoretical model results computed with the Warsaw non-linear hydrodynamics code.

\section{Observations and Data Analysis}

\begin{figure*}
\centering
\plottwo{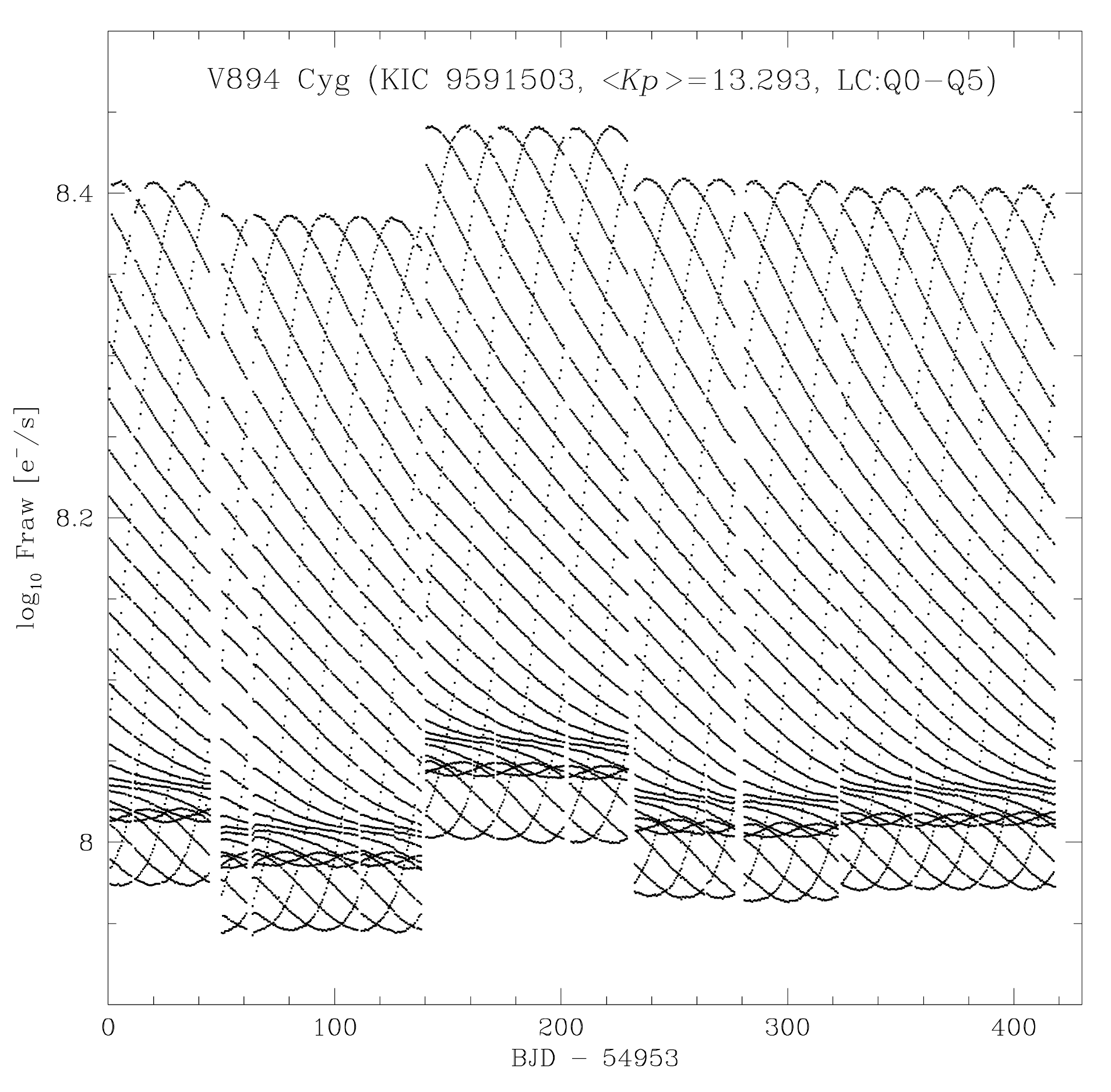}
        {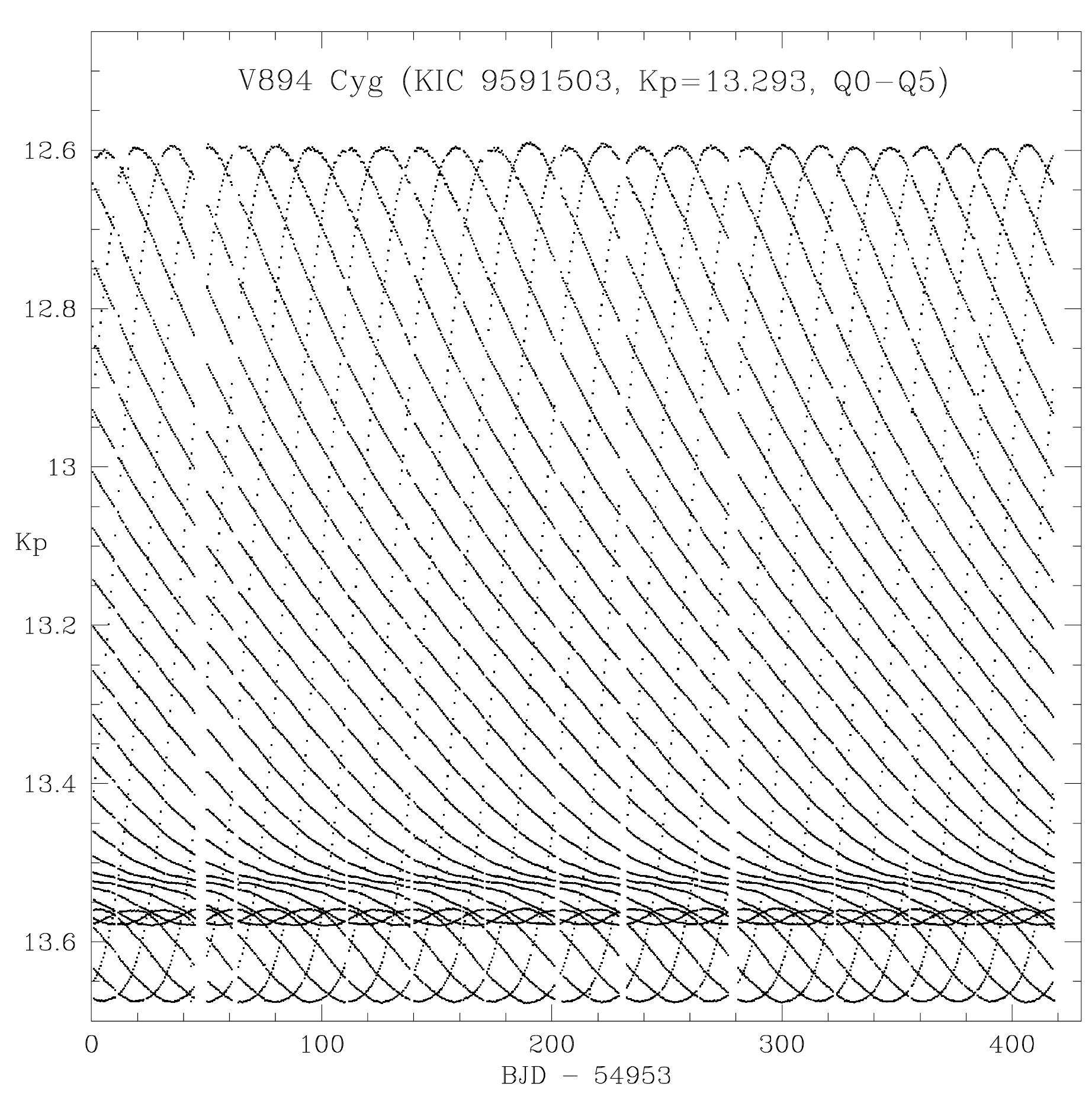}
\vskip0pt
\caption{Raw-flux (left) and $Kp$-magnitude (right) long-cadence photometry for V894~Cyg, one of the 19 non-Blazhko RR~Lyrae 
stars in the {\it Kepler} field.  The
processing of the raw-flux data consisted of normalizing the different (Q0-Q5) sensitivities, removing
(minor) trending within each quarter, stitching
together the quarters, and shifting the magnitudes to $\langle Kp \rangle = 13.293$ mag.  The times are
BJD (Barycentric Julian Date) minus 54953.0 (the first day of {\it Kepler} observations).  }
\label{V894rawproc}
\end{figure*}

For each of the 19 stars listed in {\bf Table~1} the {\it Kepler} raw-flux data 
acquired over the first 420 days of the telescope's operation (Q0-Q5) was analysed.  The mean apparent magnitudes of the
stars (from the {\it Kepler} Input Catalog) range from $\langle Kp \rangle = 12.7$ to 17.4 mag, with distances
$\sim$3-23 Kpc in the direction $(l,b) = (76.3^{\circ},13.5^{\circ})$.  
The last two columns of Table~1 identify the particular $Kepler$ data that was analyzed.

{\bf Figure~1} shows the raw-flux (left panel) and fully-transformed $Kp$-magnitude (right panel) 
as a function of time for one of the non-Blazhko stars, V894~Cyg (graphs such as these are typical).   
The flux is in units of $e^{-}$/s/cadence (log$_{10}$ scale) and 18911 
long cadence (LC) aperture photometry measurements are plotted.    
Since the 1.4-m telescope was rolled by 90$^\circ$ every three months resulting in a given
star being observed with different pixels every `quarter', it was necessary to correct 
for sensitivity variations from quarter to quarter, and for drift within each
quarter.  The $Kp$ magnitudes were derived from the fully normalized and detrended photometry. 
The scalloped pattern seen at maximum and minimum light, and the
moir\'e pattern seen everywhere, result from the pulsation period for this star, 0.571~d, not being 
evenly divisible by the LC sampling period, $\sim$30 min, giving rise to 27 observations per pulsation cycle.
The constancy of the range in the total amplitude is remarkable, as is the continuous moir\'e pattern.

The pulsation period for each star (fundamental mode) was derived from the high-precision 
aperture photometry using the PERIOD04 period-finding program (Lenz \& Breger 2005) and
a version of the CLEAN program written by Dr. Seung-Lee Kim (see Nemec, Walker \& Jeon 2009).  Typically the LC data consisted
of $\sim$18000 points, and the short cadence (SC) observations of FN~Lyr and AW~Dra (Q0 and Q5) amounted to another $\sim$150000 data points per star.   
The uncertainties in the derived periods are $\sim$1-2$\times10^7$d, depending on the particular data set 
and pulsation period, $P$.   Times of maximum light, $t_0$, accurate to $\pm$0.0005~d also were computed.  
The $P$ and $t_0$ values are summarized in columns 4 and 5 of Table~1.

\begin{figure*}
\centering
\plottwo{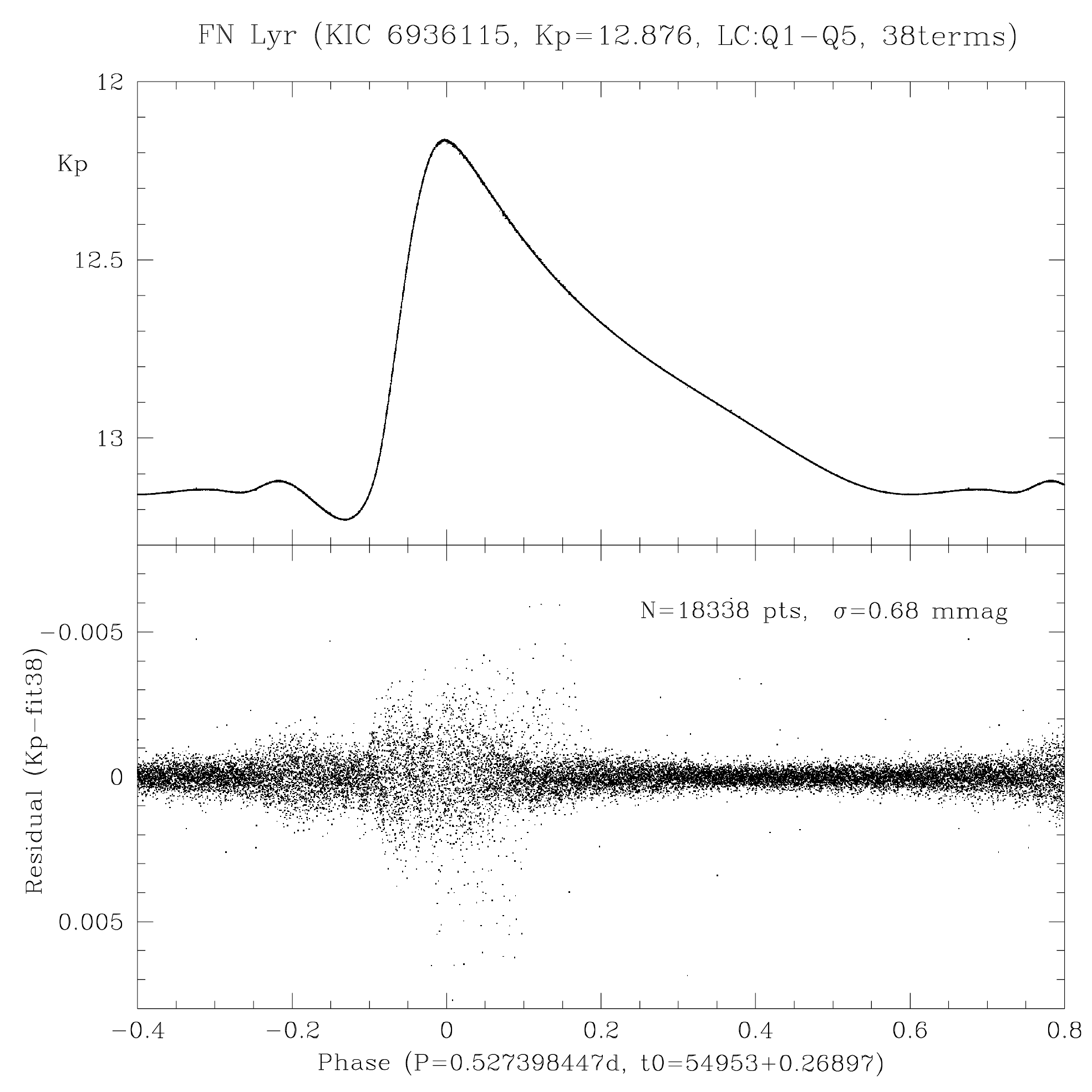}
        {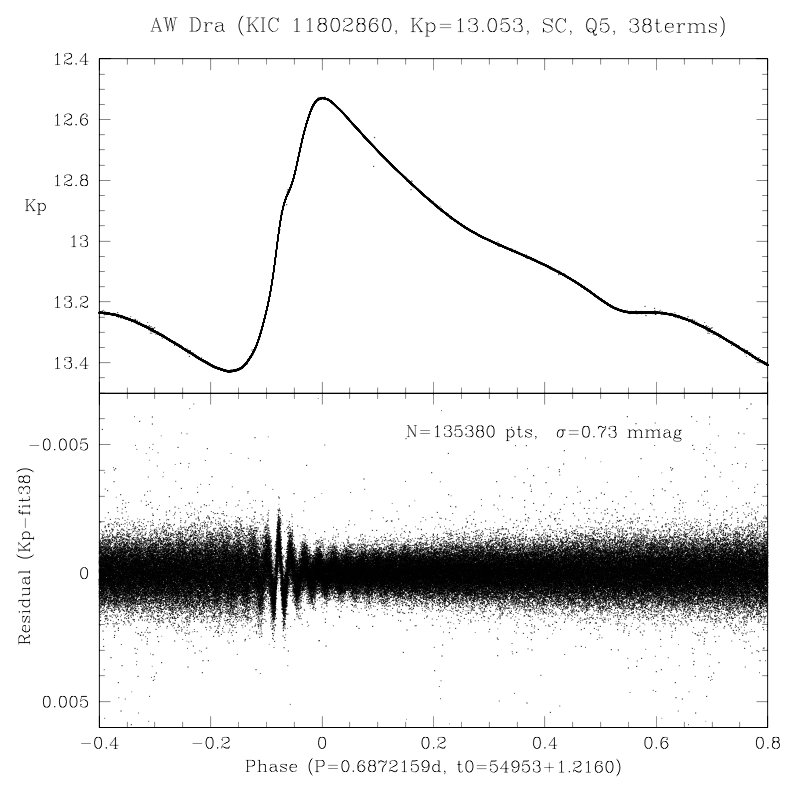}
\vskip0pt
\caption{(Left) Light curve for FN Lyr derived from Q1-Q5 long-cadence photometry;  
(Right) Light curve for AW Dra derived from short-cadence Q5 photometry.   (Lower) The residual
plots in both cases are from the 38-term Fourier fits (observed minus fitted).  The standard deviations 
of the fits in both cases are below 1 mmag and would be even smaller with more terms in
the Fourier fit. }
\label{V894rawproc}
\end{figure*}

Phased light curves for all 19 non-Blazhko stars were plotted using the derived pulsation 
periods and times of maximum light.  Two such light curves are illustrated in the top panels of {\bf Figure~2}.  The light curve
for FN~Lyr is based on 18338 LC data points acquired over $\sim$400 days in Q1-Q5, and that on the right 
is for AW~Dra and is based on 135380 SC data 
points acquired over 90 days in Q5.   When the light curves for all the
stars are plotted one clearly sees variations in the `risetimes' ({\it i.e.}, the time from minimum
to maximum light, expressed in phase units), the stars with the shortest risetimes having
the largest amplitudes and a tendency to have a bump on the rise to maximum light.

Also analyzed were new high precision ground-based $V$ photometry for three of the stars (NR~Lyr, FN~Lyr and AW~Dra),
and All Sky Automated Survey (ASAS) $V,I$ photometry (see Pigulski {\it et al.} 2009) for nine of the brightest stars.
The high-precision calculations were instrumental in establishing the $V$-$Kp$ offsets in the Fourier parameters
(needed to permit use of the Kov\'acs, Jurcsik {\it et al.} relations);  and the ASAS-North data 
(http://www.astrouw.edu.pl/asas/kepler), although not providing high-precision Fourier
parameters, were sufficient for deriving reliable loop diagrams in the HR-diagram, from which it is clear that
in every case the non-Blazhko stars are bluest when brightest (as expected).

By combining the historical data (going back more than 100 years for FN~Lyr, NQ~Lyr and AW~Dra) 
with the {\it Kepler} data we were also able to derive period change rates for some of the stars.
In particular, the periods for FN~Lyr and AW~Dra were seen to be increasing.   Unfortunately the baseline of the
{\it Kepler} observations is still too short to derive significant period variations directly.

Finally, one of the non-Blazhko stars,
KIC~7021124, was discovered to be doubly-periodic and to have properties very similar to V350~Lyr (Benk\H o {\it et al.} 2010).
Both appear to be pulsating in the fundamental and second-overtone modes simultaneously.
The period ratios are almost identical,  $P_2/P_0 = 0.59305$ for KIC~7021124, and 0.592 for V350~Lyr.  
When masses are derived for these two stars using the Petersen $P_2/P_0$ vs. $P_0$ diagram
the best agreement with the models is obtained, in both cases, for a high luminosity and a high mass:
$L/L_{\odot}=70$ and  ${\cal M}/{\cal M}_{\odot} = 0.75$.

\section {Fourier Analysis}

Fourier decomposition of the light curves was performed on the fully calibrated photometry for all the stars
by fitting the following Fourier series to the apparent magnitudes:  
\begin{equation}
m(t) = A_0 + \sum_{i=1}^F  A_i \sin  [ 2 \pi i f_0 (t - t_0)  + \phi_i ], \\
\end{equation}
where $m$($t$) is the apparent magnitude (either $Kp$ for the {\it Kepler} data, or $m_V$ for the ground based
$V$-photometry), $F$ is the number of fitted terms, $f_0$ is the (assumed single) pulsation frequency
of the star (=$P_0^{-1}$), $t$ is the observed time of the observation (BJD-54953 for the {\it Kepler} data), $t_0$ is the
time of maximum light used to phase the light curves (so  that maximum light occurs at zero phase), 
and the $A_i$ and $\phi_i$ are the amplitude and phase coefficients for the individual Fourier terms.
The Fourier calculations were made with two FORTRAN programs, one kindly provided by Dr. G\'eza Kov\'acs, and
the other by Dr. Pawel Moskalik.  The assumed pulsation periods were those derived either from the {\it Kepler}
data or from the period change rate analysis, and the $t_0$ values were calculated numerically.

The Fourier analyses were performed on various subsets of the data. 
The bottom panels of Fig.2 show residual
plots after fitting 38-term Fourier descriptions to all the data.  In both cases the standard deviations about the mean
curve are below 1 mmag, specifically $\sim$60 $\mu$magn, and the greatest differences occur where
the slope is steepest.  For the purposes of using extant Fourier correlations to derive physical
characteristics only the lower order Fourier terms are needed.  Owing to the small risetimes
as many as 100 or more terms could be needed for a normal distribution of the residuals.

\begin{figure*}
\centering
\plottwo {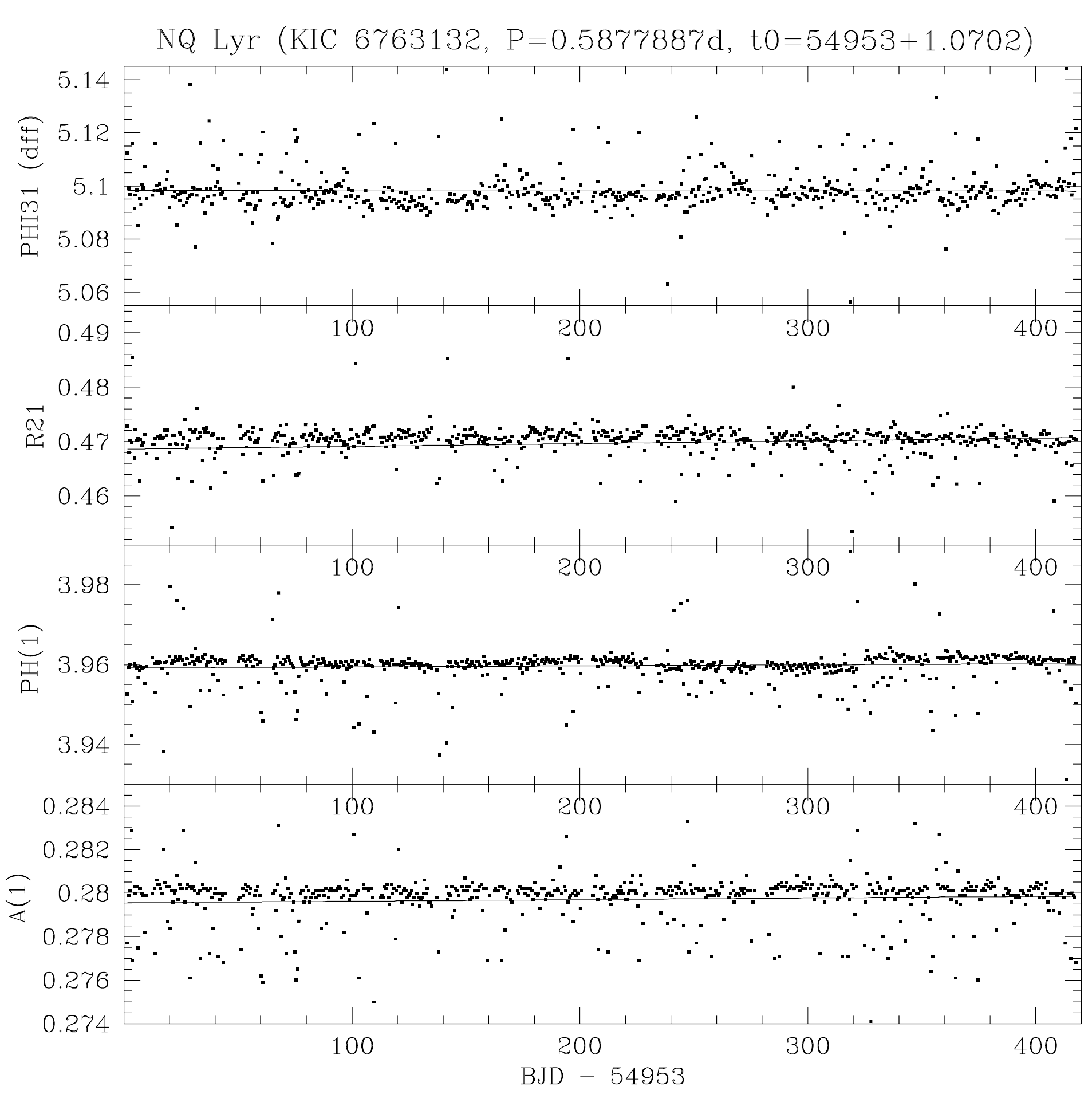} {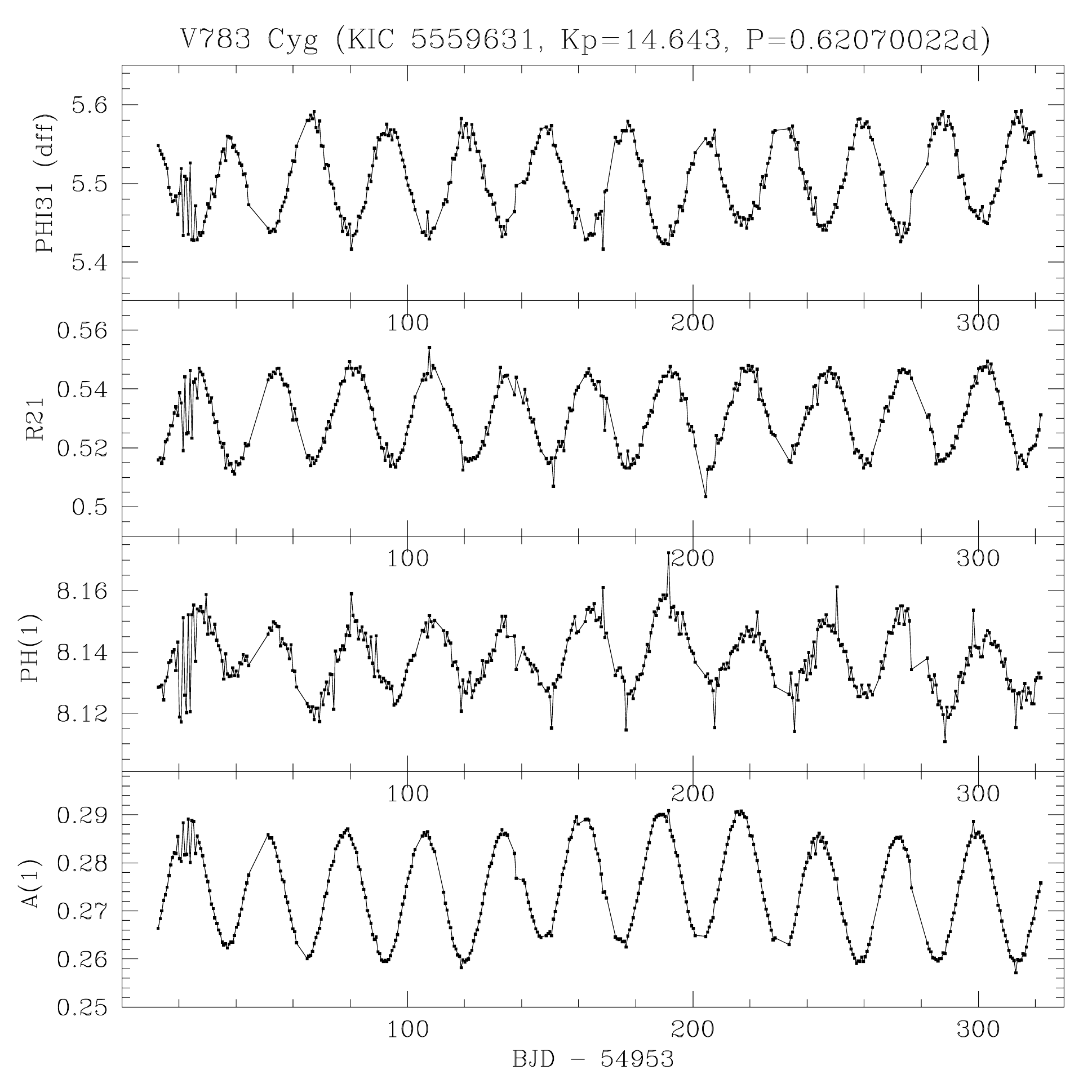}
\vskip0pt
\caption{Variations in time of the Fourier parameters PHI31, R21, PHI1S and A1 derived for the
typical non-Blazhko star NQ Lyr (left panel) and for the known low-amplitude
Blazhko star V783~Cyg (right panel).  Each point corresponds to the variable of interest
as determined from a single pulsation cycle. }
\label{nonBL_vs_BL}
\end{figure*}

Another way of looking at the data was to calculate Fourier coefficients and parameters for each pulsation cycle.
For the long cadence (30 min) data there were $\sim$24 observations per bin for an RRab star
with a period of 12 hours, and fewer (more) observations per cycle for stars with shorter (longer) periods.   
None of the 19 stars exhibits the recently discovered "period doubling" effect, thus confirming the
result found earlier by Szab\'o {\it et al.} (2010).  
For the Fourier calculations 
`direct Fourier fitting' (dff) rather than `template Fourier fitting' (tff) methods were used (see Kov\'acs \& Kupi 2007),
and for each star the resulting time series were plotted for four Fourier parameters: 
$A_1$, the first term in the Fourier series (see right panel of Fig.~2);  $\phi_1$, the phase of the first term; 
$R_{21}$, the amplitude ratio $A_2/A_1$; and $\phi_{31}^s$, the Fourier parameter found by 
Simon, Kov\'acs and others to be one of the most significant variables for deriving physical characteristics.
Typical time series are illustrated in {\bf Figure~3}, on the left for the unmodulated RRab star NQ~Lyr having 
intermediate amplitude ($A_1$=0.279 mag) and intermediate brightness, $\langle Kp \rangle =13.075$~mag, and 
on the right for V783~Cyg, a low-amplitude 
Blazhko star with the shortest Blazhko period ($P_B = 27.7\pm0.04$~d)
among the stars in the sample. 
For NQ~Lyr linear trend lines were fit to each time series and in almost every case the slope is
zero to within the systematic and random uncertainties.  The mean values for the four parameters are
as follows:  
$\phi_{31}^s = 5.0958 \pm 0.0023$, with residual standard error, $\sigma =0.002$;
$R_{21} = 0.4710 \pm 0.0006$, with $\sigma = 0.0006$;
$\phi_1^s = 3.961 \pm 0.001$, with $\sigma = 0.001$; and
$A_1 = 0.28016 \pm 0.00014$ mag, with $\sigma = 0.0001$.
Since the other non-Blazhko stars show random variations of the Fourier parameters similar to those shown
here for NQ~Lyr these means and errors provide a measure of the typical uncertainties.       
Inspection of the NQ~Lyr time series in Fig.~3 shows that the stability over the $\sim$420~d interval (Q0-Q5) is exceptional, 
not only for each Fourier parameter but for the ensemble of parameters.  To better illustrate the remarkable
stationarity of the light curves for all 19 sample stars a set of `animated gif' light curves was 
prepared and these are available in the electronic version of the MNRAS paper.

In {\bf Table~2} some of the resulting Fourier coefficients and
parameters are summarized. 
Column 2 gives the standard deviation about the mean light curve (see Fig.2), most of which are below
one mmag.  Column 3 contains
the first term in the Fourier series, which correlates extremely well with the total amplitude. 
Columns 4 and 5 contain amplitude ratios (defined above) and columns 6 and 7 contain phase
parameters (see Simon 1988 and references therein).

\begin{deluxetable*}{lccccccc}
\tabletypesize{\small}
\tablecaption{Selected Fourier Parameters and Derived Metallicities for the {\it Kepler} non-Blazhko ab-type RR Lyrae stars}
\tablewidth{0pt}
\tablehead{ \\ \colhead{Star} & \colhead{ $\sigma$ }  &  \colhead{ $A_1$ } & 
   \colhead{  $R_{\rm 21}$  }  & \colhead{ $R_{\rm 31}$   }  & \colhead{ $\phi_{\rm 21}^s$  }  & 
    \colhead { $\phi_{\rm 31}^s$  } & \colhead{ [Fe/H]$_{\rm ZW}$   } \\ 
\colhead{} & \colhead{[mmag]}  & \colhead{[mag]} & \colhead{} & \colhead{} & \colhead{[rad]} & \colhead{[rad]} & \colhead{ } \\
} 
\startdata
NR~Lyr            & 0.69   & 0.266 & 0.456 & 0.352 & 2.416 & 5.115 & -2.34 \\ 
V715~Cyg          & 1.74   & 0.338 & 0.479 & 0.358 & 2.314 & 4.901 & -1.44 \\                      
V782~Cyg          & 0.79   & 0.190 & 0.488 & 0.279 & 2.777 & 5.808 & -0.47 \\
V784~Cyg          & 0.96   & 0.234 & 0.487 & 0.253 & 2.904 & 6.084 & -0.14 \\
KIC~6100702       & 0.66   & 0.209 & 0.493 & 0.279 & 2.743 & 5.747 & -0.35 \\
NQ~Lyr            & 0.65   & 0.280 & 0.471 & 0.356 & 2.389 & 5.096 & -1.83 \\
FN~Lyr            & 0.64   & 0.380 & 0.444 & 0.350 & 2.322 & 4.818 & -1.90 \\
KIC~7021124       & 1.10   & 0.283 & 0.512 & 0.351 & 2.372 & 5.060 & -2.08 \\
KIC~7030715       & 0.71   & 0.231 & 0.494 & 0.303 & 2.683 & 5.606 & -1.66 \\
V349~Lyr          & 3.24   & 0.346 & 0.450 & 0.352 & 2.328 & 4.845 & -1.73 \\
V368~Lyr          & 1.57   & 0.405 & 0.464 & 0.341 & 2.272 & 4.784 & -1.53 \\
V1510~Cyg         & 0.82   & 0.345 & 0.473 & 0.355 & 2.389 & 5.068 & -1.83 \\
V346~Lyr          & 2.83   & 0.330 & 0.473 & 0.352 & 2.372 & 5.060 & -1.83 \\
V350~Lyr          & 1.67   & 0.340 & 0.485 & 0.342 & 2.389 & 5.124 & -1.84 \\
V894~Cyg          & 0.91   & 0.377 & 0.490 & 0.338 & 2.364 & 5.067 & -1.79 \\
V2470~Cyg         & 0.79   & 0.220 & 0.488 & 0.282 & 2.745 & 5.737 & -0.71 \\
V1107~Cyg         & 0.99   & 0.280 & 0.495 & 0.350 & 2.421 & 5.196 & -1.56 \\
V838~Cyg          & 1.22   & 0.393 & 0.465 & 0.349 & 2.300 & 4.853 & -1.56 \\
AW~Dra            & 0.55   & 0.306 & 0.525 & 0.346 & 2.730 & 5.560 & -1.74 \\
\enddata  
\end{deluxetable*}

It is of interest to compare the derived Fourier parameters 
for the 19 non-Blazhko stars in the {\it Kepler} field (derived using the $Kp$ photometry
but transformed to $V$ values using the offsets) with parameters derived for Galactic globular
clusters and for RR~Lyrae stars in the Large Magellanic Cloud.

\begin{figure*}
\centering
\plottwo{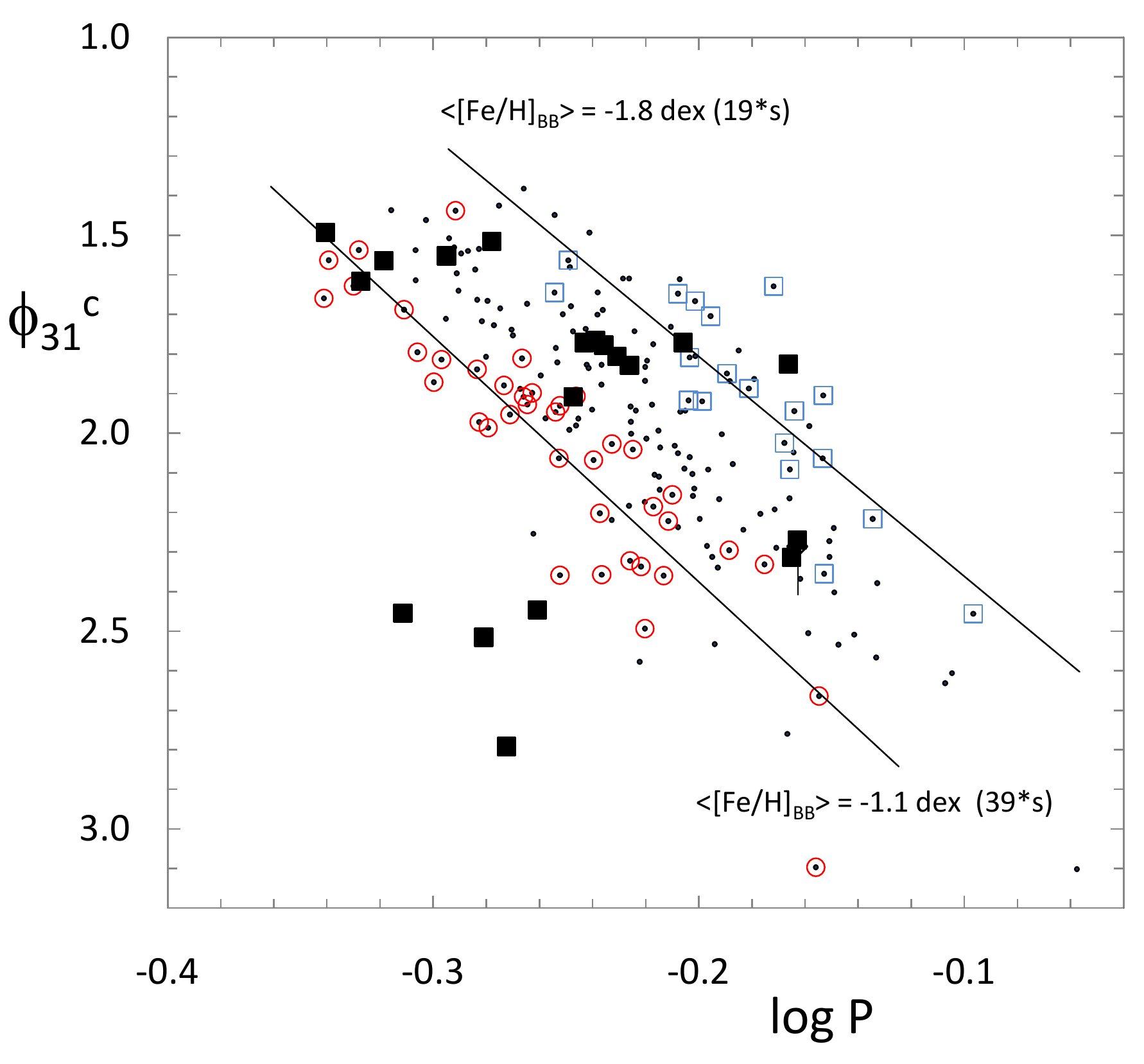}  {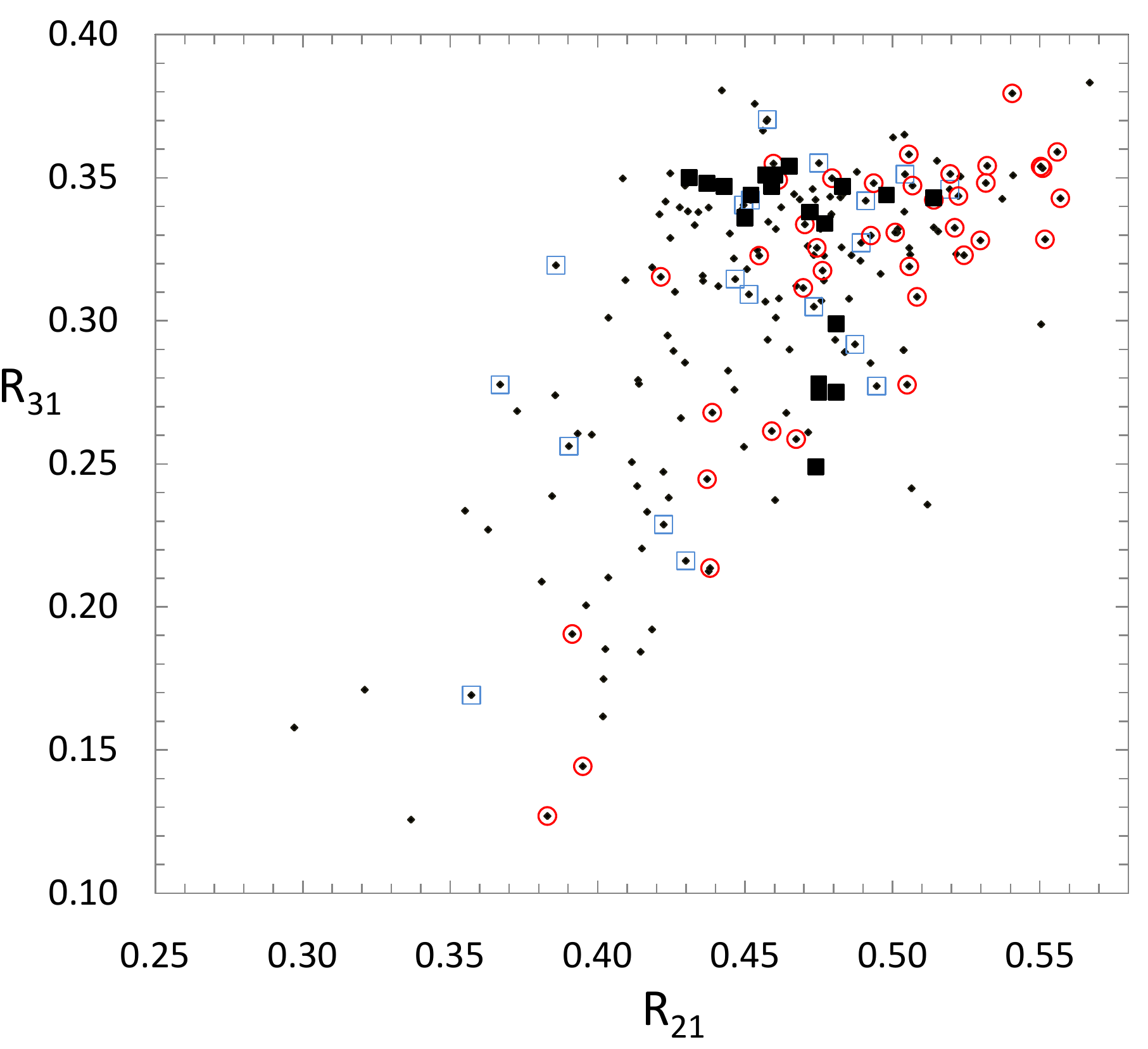} 
\plottwo{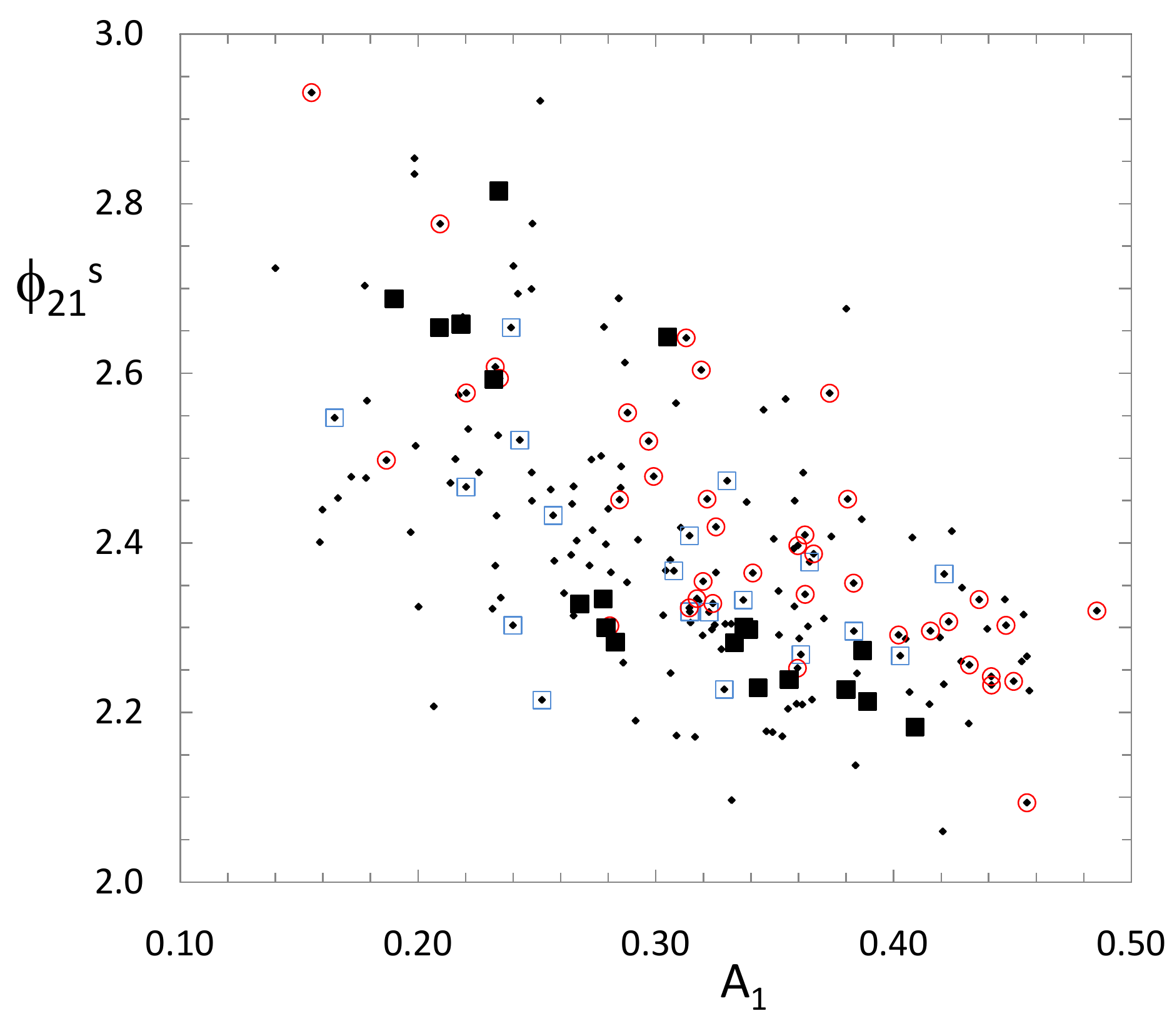}  {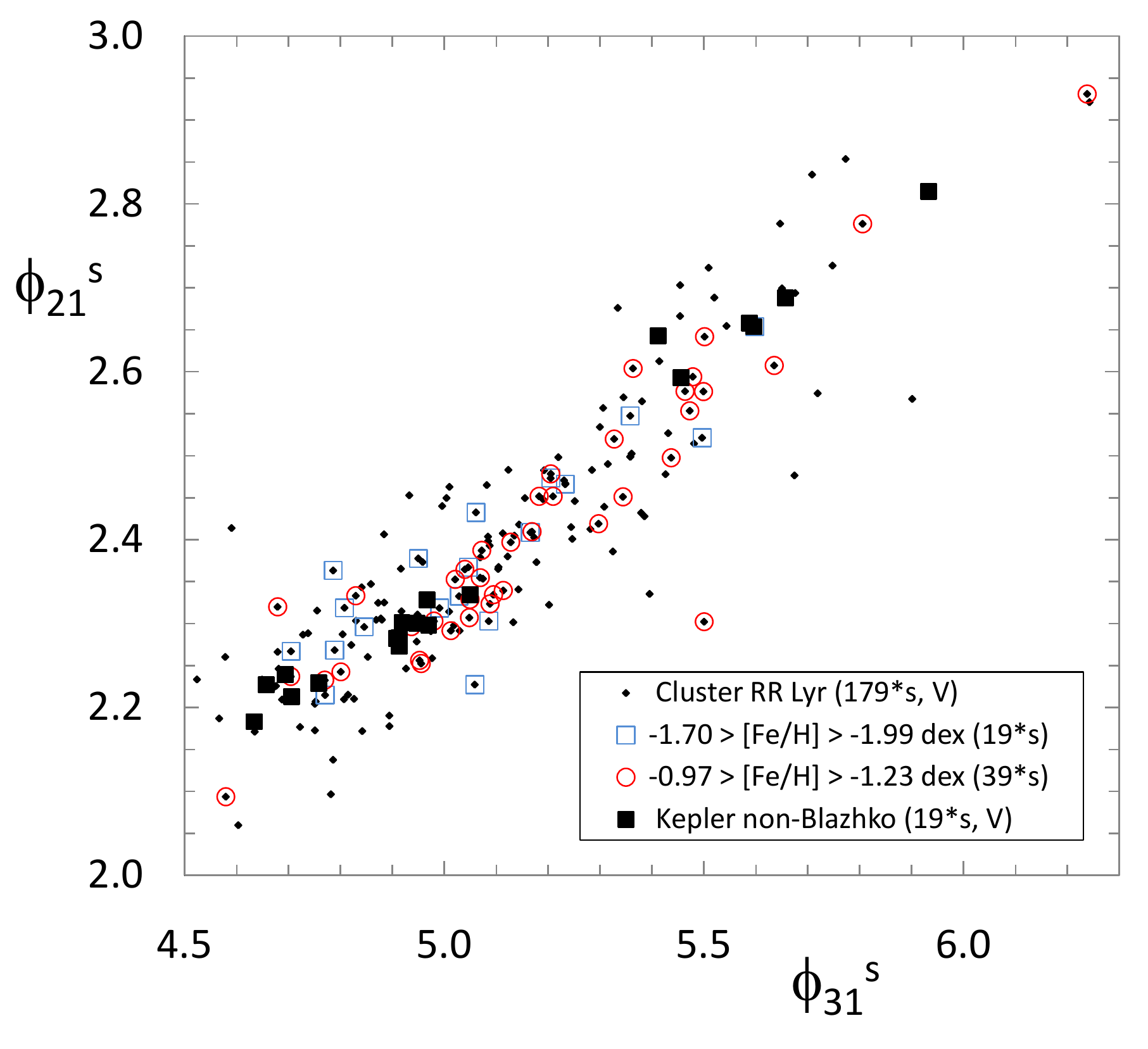}
\vskip0pt
\caption{Four panels comparing the Fourier parameters (transformed to the $V$-band) for 
the {\it Kepler} non-Blazhko stars (large black squares) and 
for the 177 RR~Lyr stars located in several Galactic and LMC globular clusters (small black dots).
The cluster RR~Lyr stars (from Kov\'acs \& Walker 2001) are in the 
globular clusters M2 (12 stars), M4 (4), M5 (12), M55 (4), M68 (5), M92 (5), NGC~1851 (11), 
NGC~5466 (6), NGC~6362 (12), NGC~6981 (20), IC~4499 (49), Ruprecht~106 (12), NGC~1466 (8), Reticulum (8) and NGC~1841 (9). 
The points for the 19 most metal-poor globular cluster stars are surrounded 
by blue squares, and the points for the 39 most metal-rich globular cluster stars are circled
in red.  }
\label{4panel}
\end{figure*}

{\bf Figure 4} compares the non-Blazhko stars with 177 RR~Lyr stars in several well-studied globular clusters (from Kov\'acs \& Walker 2001). 
The same symbols and colour coding are used in all four panels.  
The {\bf upper-left} panel is analogous to the $P$-$A_{\rm tot}$ diagram (see Fig.6 below) and was used to derive [Fe/H] values for the {\it Kepler} stars.  
The two diagonal lines represent mean relations when the globular cluster data (small black dots) are sorted into two [Fe/H] bins (the `BB' subscript refers to the 
Butler-Blanco system, which approximates the Carretta-Gratton system): 
a metal-poor bin consisting of 19 stars (surrounded by blue squares) with metallicities between $-1.70$ and $-1.99$ dex and average $-1.8$ dex (BB-scale);  and 
an intermediate-metallicity bin consisting of 39 stars (circled in red) with [Fe/H]$_{\rm BB}$ between $-0.97$ and $-1.23$ dex and average $-1.1$ dex.
The equations of the lines are: $\phi_{31}^c = 5.556 \thinspace {\rm log}P + 0.2920$ (upper metal-poor bin)
and  $\phi_{31}^c = 6.200 \thinspace {\rm log}P + 3.615$ (lower intermediate metallicity bin).
Four of the {\it Kepler} non-Blazhko stars are clearly richer than [Fe/H]$_{\rm BB}$=$-1.1$ dex, 
while the remainder are apparently more metal poor.  
The {\bf upper-right} panel of Fig.~4 shows that the {\it Kepler} stars appear to be drawn from a distribution similar to
that of the globular clusters.  In particular, the {\it Kepler} stars with low $R_{31}$ values are not unusual,
except possibly that they all have relatively high $R_{21}$ values.
Since the globular cluster stars of higher metallicity (red open circles) all tend to reside on the right side of the 
diagram this separation is probably a metallicity effect,  supporting our conclusion that 
V784~Cyg, V782~Cyg, KIC~6100702 and V2470~Cyg are metal rich.
Likewise, the {\bf lower-left} panel shows that the majority of the {\it Kepler} stars do not differ 
from the stars in globular clusters.
Note too that there is very little metallicity discrimination in this plane.
The two stars located at the extreme upper edge of the envelope of the globular cluster distribution are AW~Dra and V784~Cyg.
Finally, the {\bf lower-right} panel shows close agreement between the phase parameters of the {\it Kepler} and globular 
cluster RR~Lyr stars, which supports the conclusion  drawn earlier (lower-right panel in Fig.~4) that
there is a strong approximately-linear correlation between $\phi_{21}^s$ and $\phi_{31}^s$.
The  diagram also shows very little  dependence on metallicity.

{\bf Figure~5} compares the  $R_{21}$,  $R_{31}$,  $\phi_{21}^c$ and $\phi_{31}^c$ 
Fourier parameters for the 19 {\it Kepler} non-Blazhko RR~Lyr stars 
with those for the field RR~Lyr stars in the central regions of the LMC.   
The LMC data are from the massive OGLE-III catalog of variable stars by Soszynski {\it et al.} (2009), which comprises 
almost 25000 RR~Lyr stars.  All the parameters are $V$-band values, the LMC values having been
transformed from $I$ to $V$ using the transformation equations given by 
Morgan, Simet \& Bargenquast (1998), and the values for the  
{\it Kepler} RR~Lyr stars transformed from $Kp$ to $V$ using the well-determined offsets from diagrams such 
as Fig.6 below.   The {\bf upper-left} panel shows that the $R_{21}$ values for the {\it Kepler} stars are typical for ab-type RR~Lyr stars and
they all lie in the narrow range 0.45 to 0.51.  The period ranges of the RRab stars also are similar.   
Where the LMC stars differ from
the {\it Kepler} stars is in the much larger $R_{21}$ range; in particular, none of the {\it Kepler}
stars are among the LMC stars in the `dropdown' at ${\rm log}P \sim -0.19$.  It is suspected that the
majority of these are RRab Blazhko variables.    
In the {\bf upper-right} panel of Fig.~5 the $R_{21}$ and $R_{31}$ amplitude ratios are compared. 
This graph puts into perspective the small ranges of $R_{21}$ and $R_{31}$ seen for the {\it Kepler} non-Blazhko 
RRab stars, both relative to that for the LMC RRab stars, and relative to the ranges seen for  
other types of RR~Lyr stars (RRc, RRd and RRe). 
The upper left panel of Fig.~4 has already shown that the {\it Kepler} stars sort into 
four metal-rich and 15 metal-poor stars, and that the metal-poor stars are similar to the 
RR~Lyr stars found in Galactic globular clusters (both OoI and OoII types);  however,
the metal-rich stars have no counterparts, at least among the 
Kov\'acs \& Walker (2001) sample of globular cluster RR~Lyr stars.  In the {\bf bottom} panels of Fig.~5 (both of which
are metallicity diagnostics) we see that both the metal-rich {\it Kepler} RR~Lyr stars (the clump of four stars at log$P\sim-0.3$ and $\phi_{21}^c \sim 4.3$)
and the metal-poor {\it Kepler} RR Lyr stars (the majority with lower $\phi_{21}^c$ values) 
have counterparts in the LMC, and that the metal-rich stars appear to have relatively
long periods for metal-rich RRab stars ({\it i.e.}, the subgroup of RRab stars with shorter periods at a given
$\phi_{31}^c$).

\begin{figure*}
\centering
\vskip0pt
\plottwo {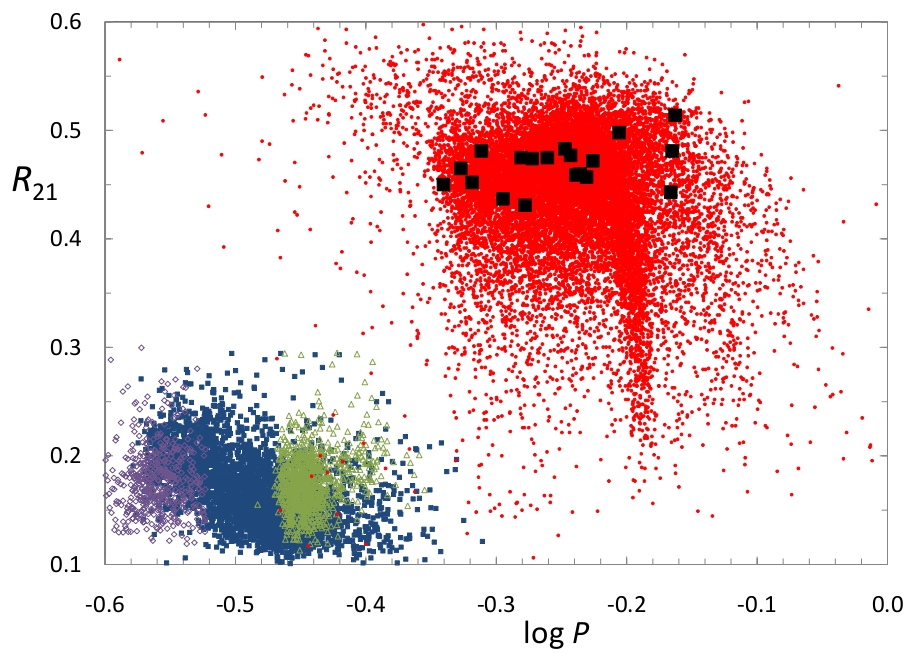} {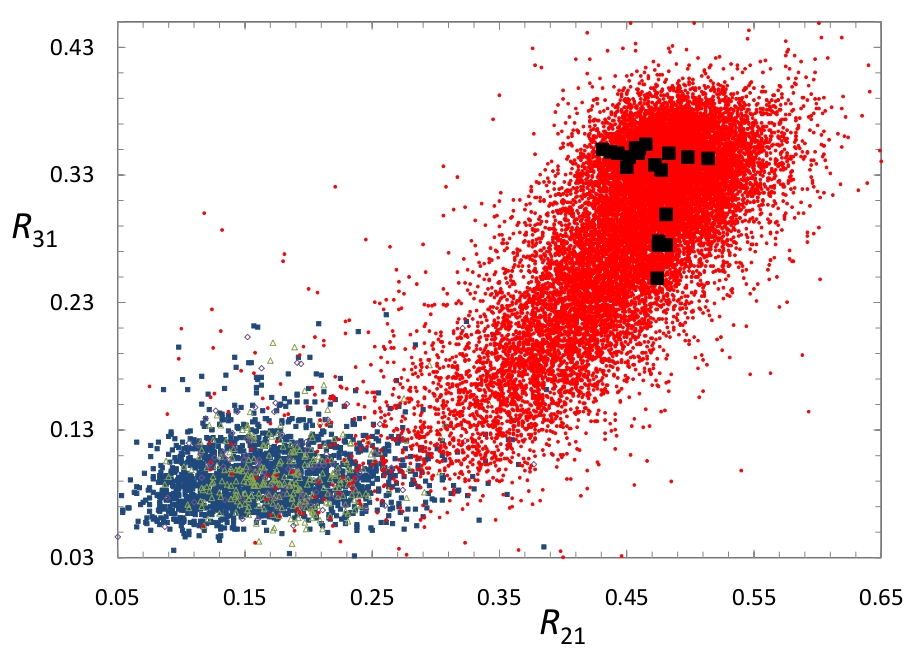} 
\plottwo {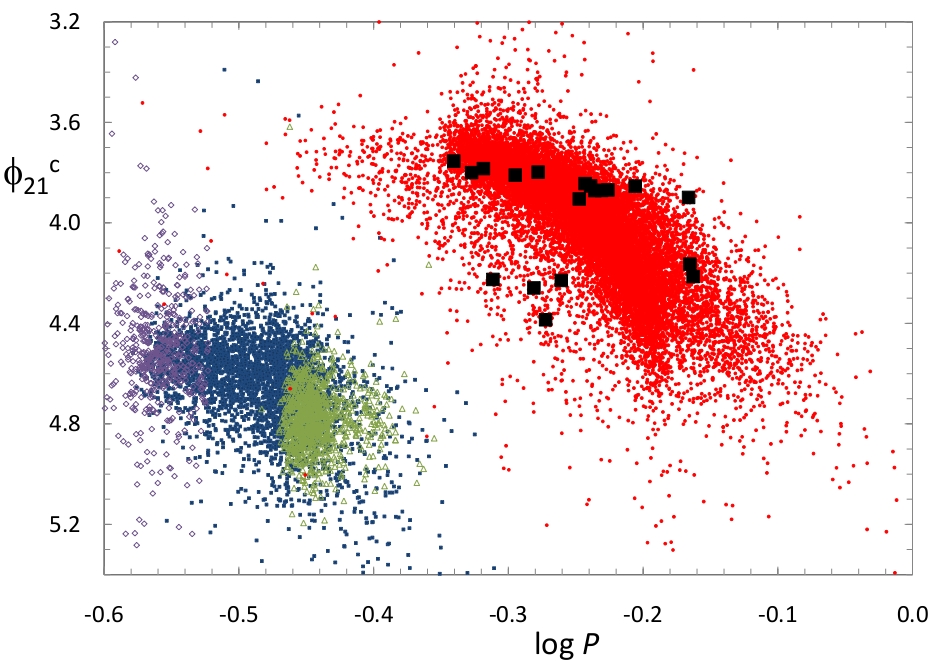} {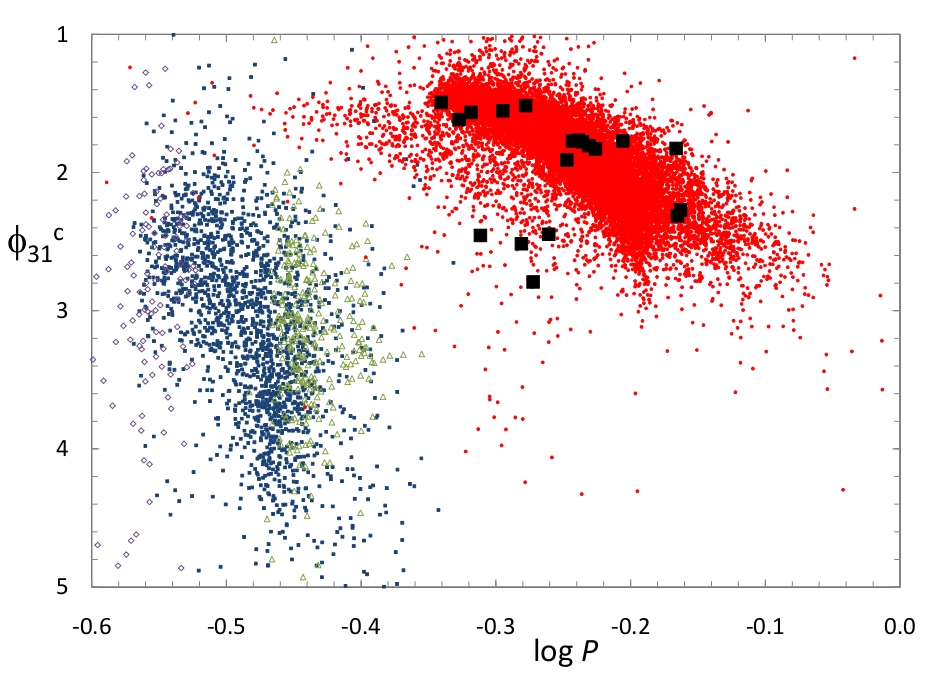}
\caption{Four panels comparing Fourier parameters ($V$-band) for the {\it Kepler} 
non-Blazhko ab-type RR~Lyr stars (large black squares) and for
24905 RR~Lyr stars in the central regions of the Large Magellanic Cloud.  
The LMC data are 
from the Soszynski {\it et al.} (2009) OGLE-III study, with colour and symbol coding as follows: 
red filled dots (17693 RRab stars);  blue filled squares (4957 RRc stars);  green open triangles (986 RRd stars), 
and purple open diamonds (1269 RRe stars). 
Because of the limited $x$- and $y$-ranges not all of the LMC RR~Lyr stars are represented in the graphs.  }
\label{OGLE-III}
\end{figure*}

\section {Physical Characteristics based on Fourier Parameters}

Several different physical characteristics were established using equations 
derived by Jurcsik (1998, hereafter J98), Kov\'acs \& Walker (2001), and Sandage (2004, 2006).
Since these equations were originally established for $V$ data, and we have only $Kp$ data, 
it was necessary first to establish $Kp$ versions of the equations.  The specific equations 
used to derive the physical quantities, and the detailed tables given in 
Nemec {\it et al.} (2011) are not repeated here, but the 
results of their application are summarized in several graphs and discussed below.

{\bf Figure~6} shows four different metallicity diagnostic diagrams.  The upper-left panel is the classical 
period-amplitude diagram, often called the `Bailey $P$-$A$ diagram' but coined the `Oosterhoff-Arp-Preston $P$-$A$ diagram' 
by Sandage (2004, hereafter S04).
At a given total-amplitude lower [Fe/H] stars tend to have longer periods.  Spectroscopically calibrated
diagonal lines, such as that shown in the lower right panel for the Messier~3 RRab stars ($\phi_{31}^c = 3.124 + 5.128 {\rm log} P$), 
can be constructed, from which offsets give [Fe/H] values. 
Since the pattern of the stars seen in all four panels is very similar all four variables can,
in principle, be used to derive [Fe/H] values.     
Four different estimates of [Fe/H] were calculated by Nemec {\it et al.} (2011), 
based on the $\phi_{31}^s$ relation of J98 (her equation~1), the $\phi_{31}^c$ relation of S04 (his equation~3), the
$A_{\rm tot}$ relation of S04 (his equation~6), and the `risetime' relation of S04 (his equation~7).
The $\phi_{\rm 31}$ relations seem to be the most reliable and the [Fe/H] values based on the J98 formula,
after conversion to the Zinn-West scale, are given in the last column of Table~2.
Recently acquired CFHT 3.6-m ESPaDOnS spectra confirm that KIC~6100702 is indeed metal-rich, at [Fe/H] = -0.18$\pm$0.06 dex, 
while AW~Dra is of intermediate metallicity, with [Fe/H] = -1.33$\pm$0.08 (Nemec, Ripepi, Chadid {\it et al.} 2011, in preparation).

\begin{figure*}
\centering
\plottwo {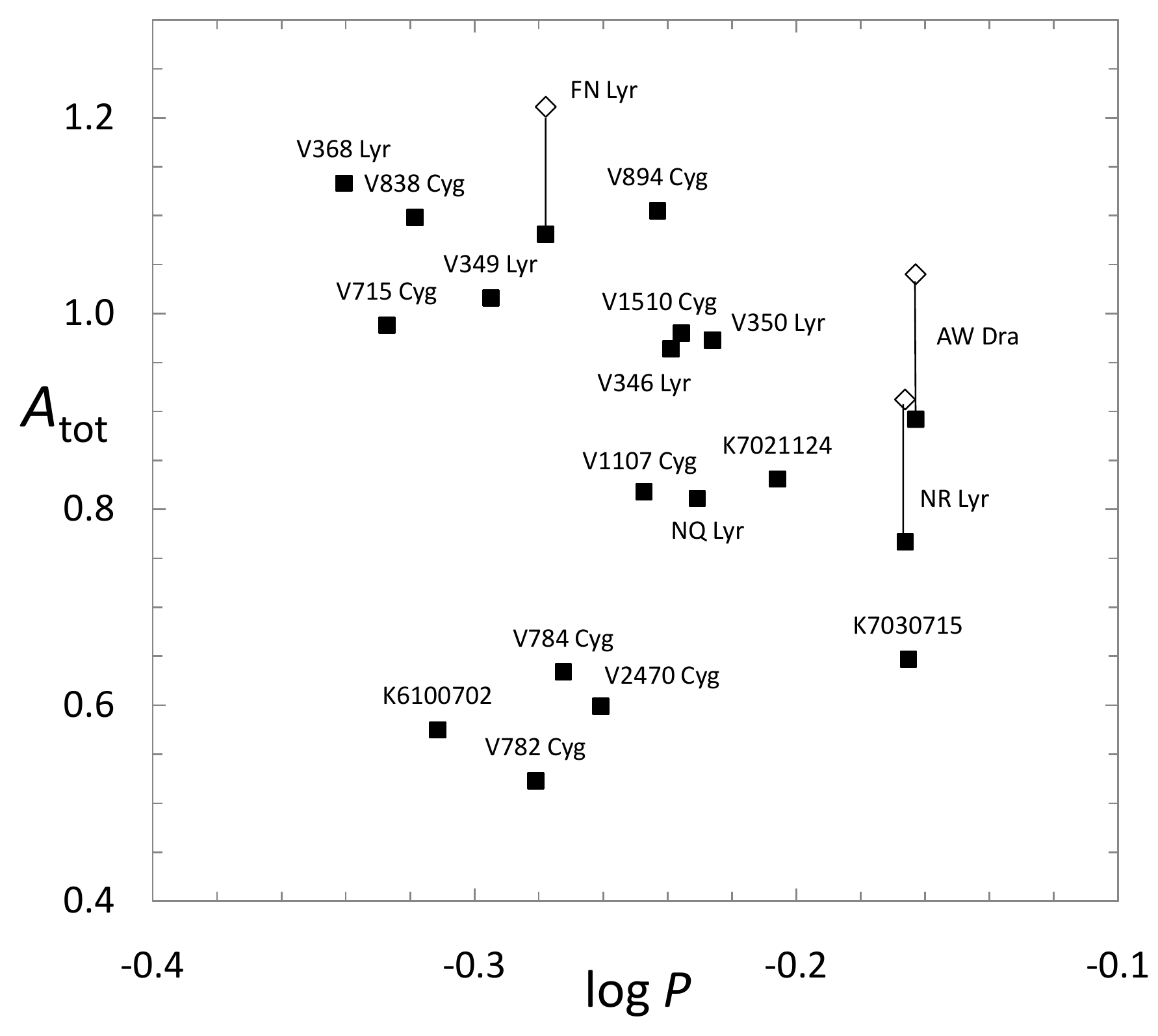}  {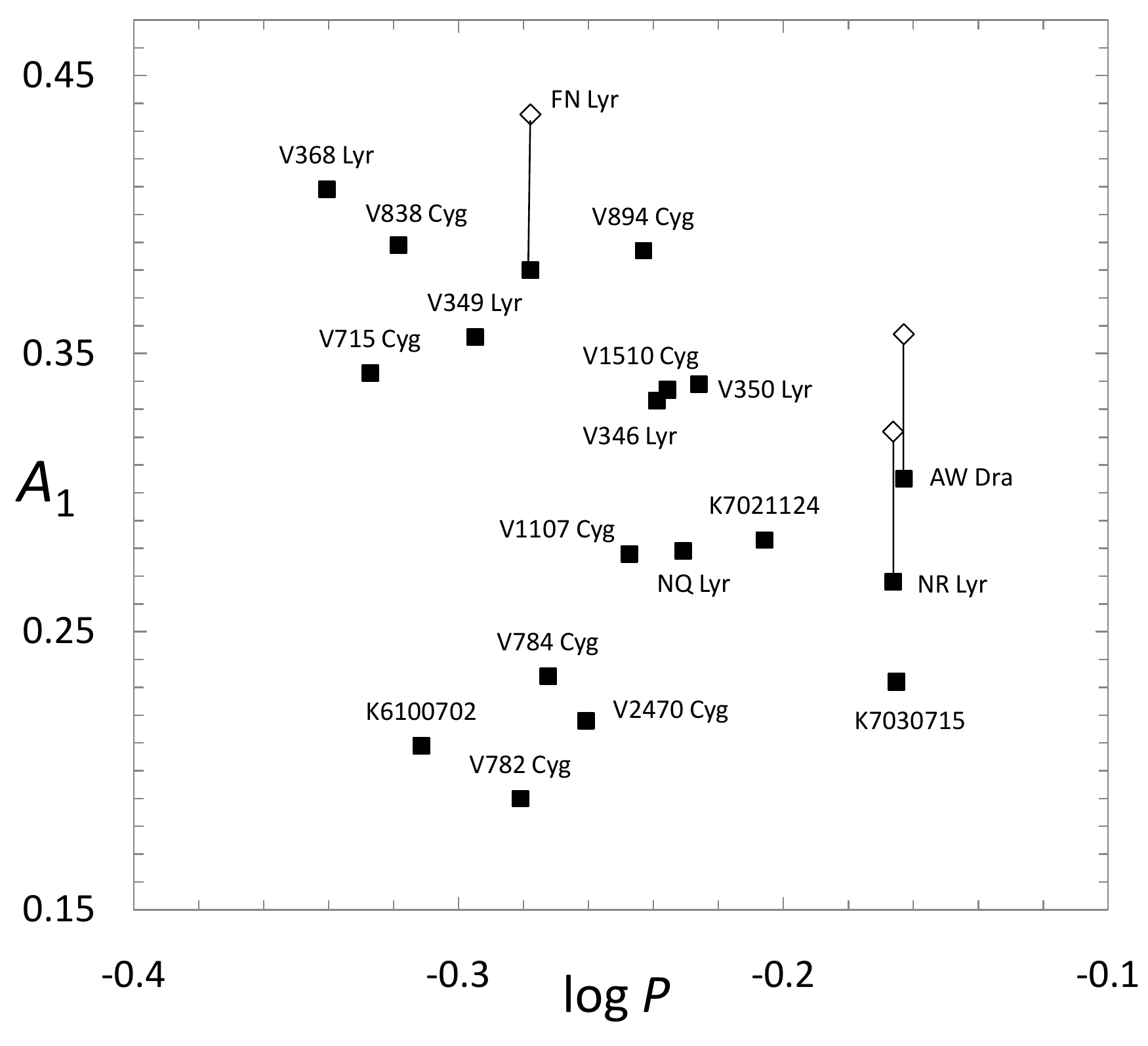} 
\plottwo {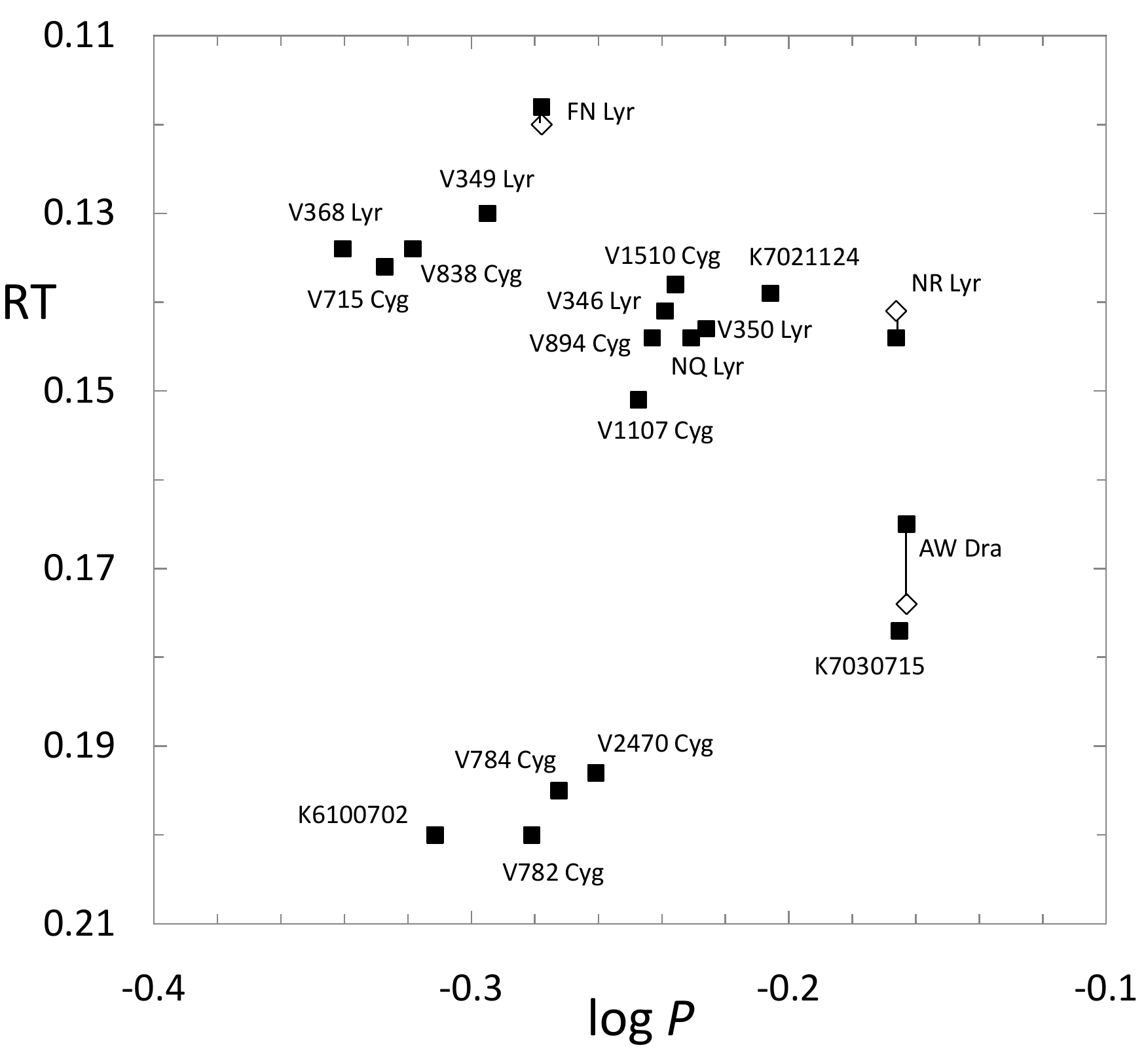} {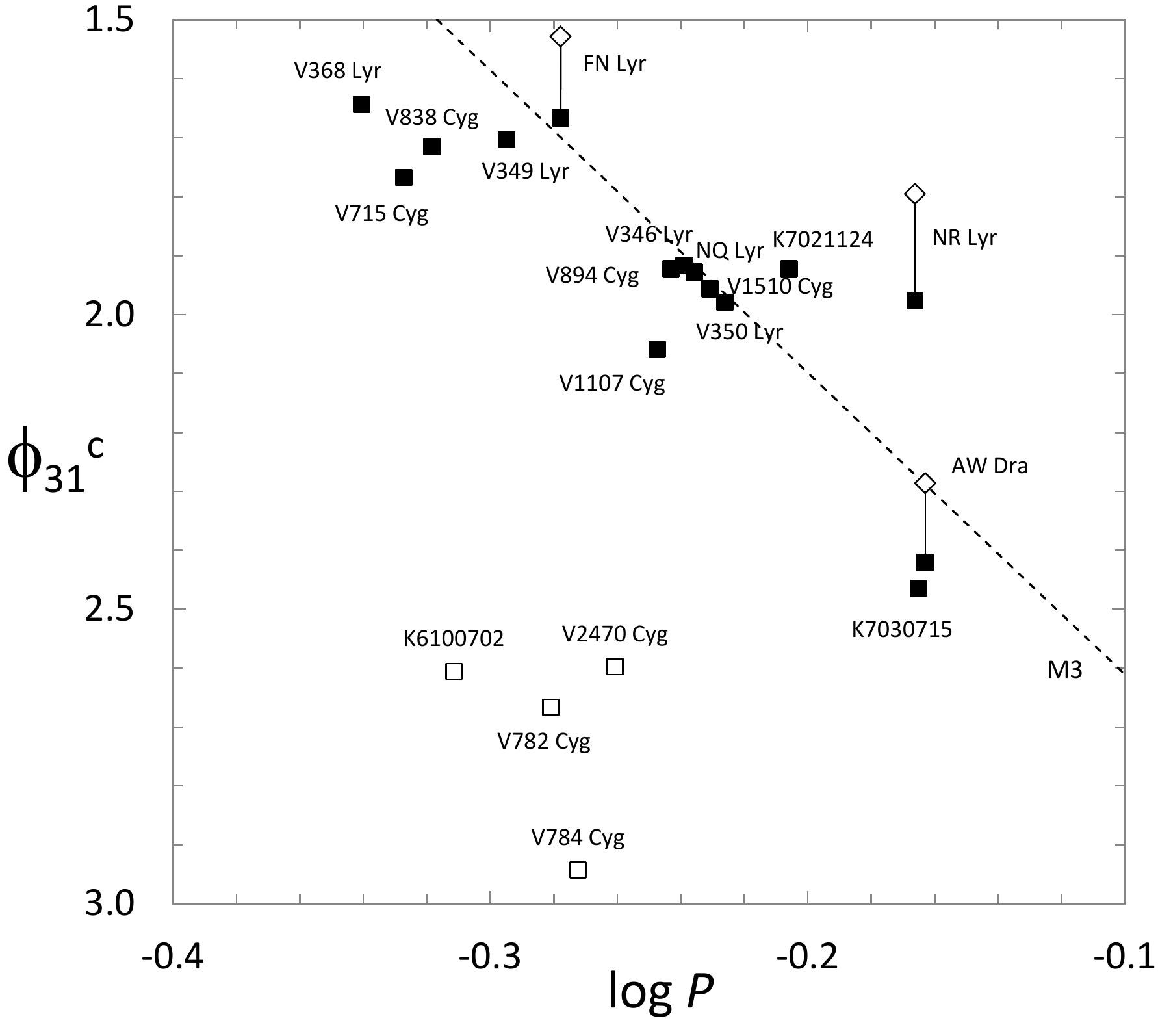}
\vskip0pt
\caption{Period-amplitude and three other metallicity 
diagnostic diagrams (all with log~$P$ along the abscissa) for the {\it Kepler} non-Blazhko RRab stars.
AW~Dra, FN~Lyr and NR~Lyr are plotted twice in each panel: the {\it Kepler} points are shown as black squares, 
the $V$-photometry points are shown as open diamonds, and the two are connected by vertical lines. 
The consistency of the $Kp$-$V$ 
offsets seen here and elsewhere justifies using the J98 and Sandage relations to calculate
physical characteristics.  In 
the lower-right panel the four metal rich stars
have been plotted as open-squares, and the dashed diagonal line is the observed $V$-relation for the M3 RR~Lyrae stars (from
Cacciari {\it et al.} 2005). }
\label{FeHDiag}
\end{figure*}

\begin{figure*}
\centering
\plottwo {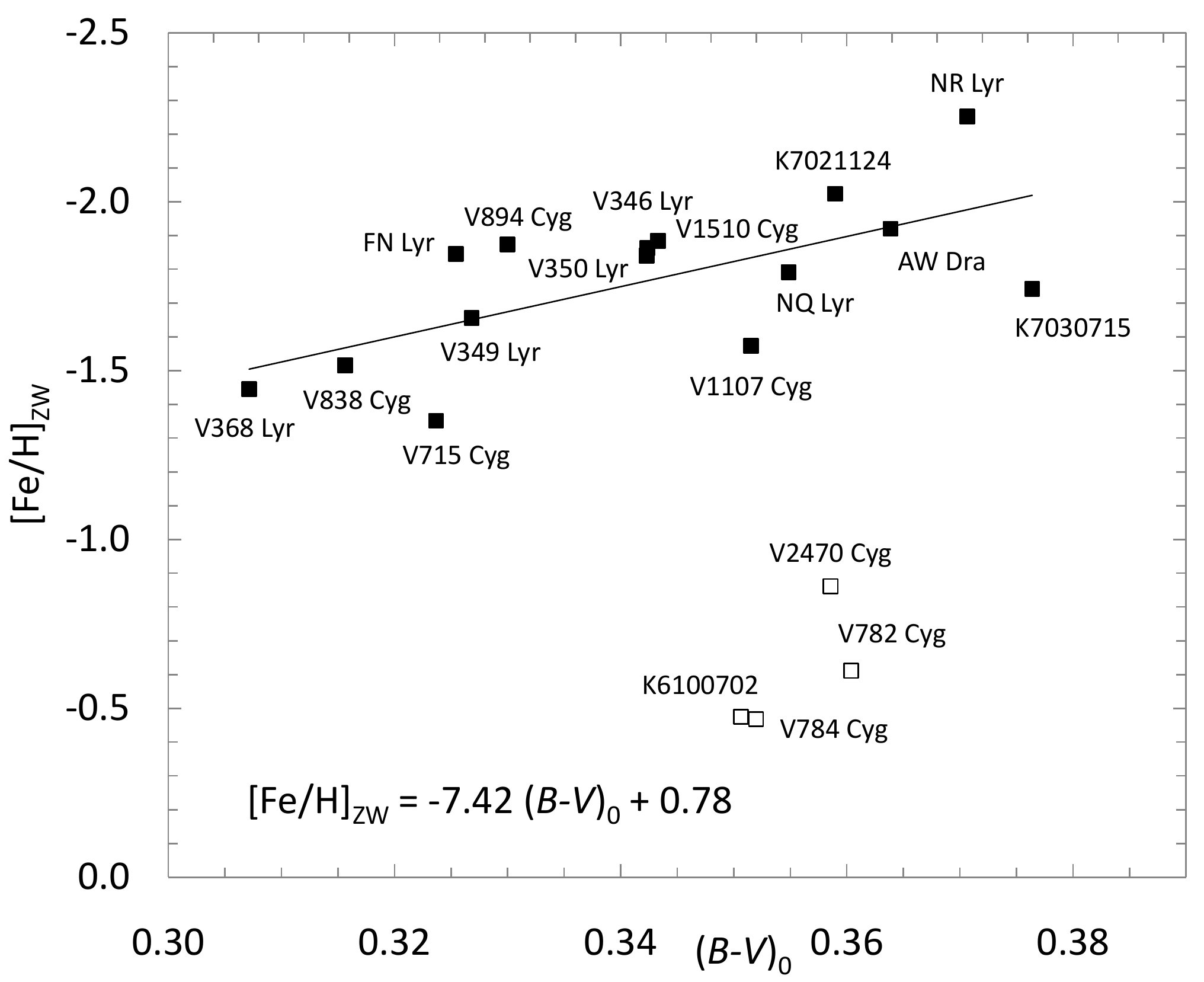} {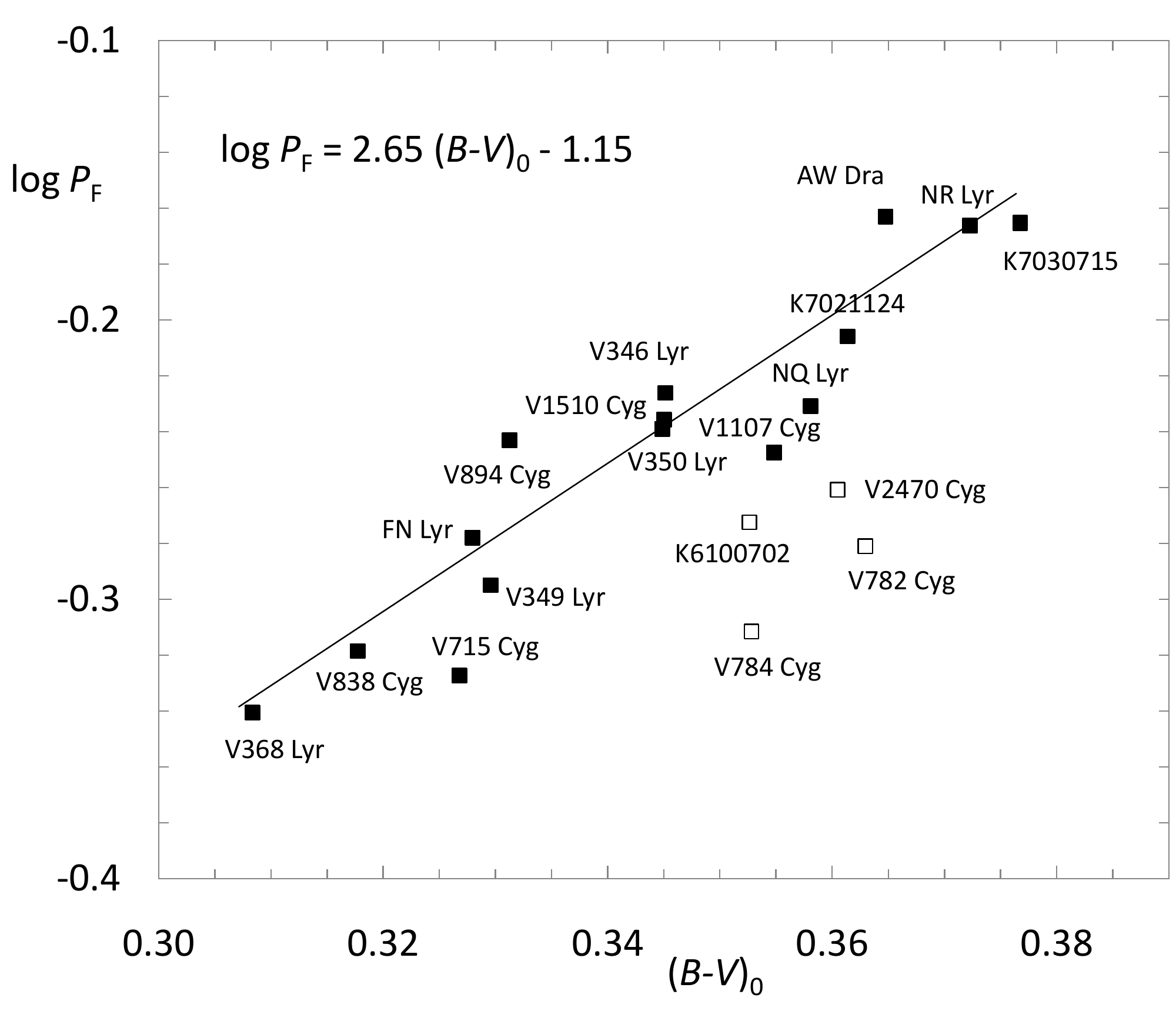} 
\plottwo {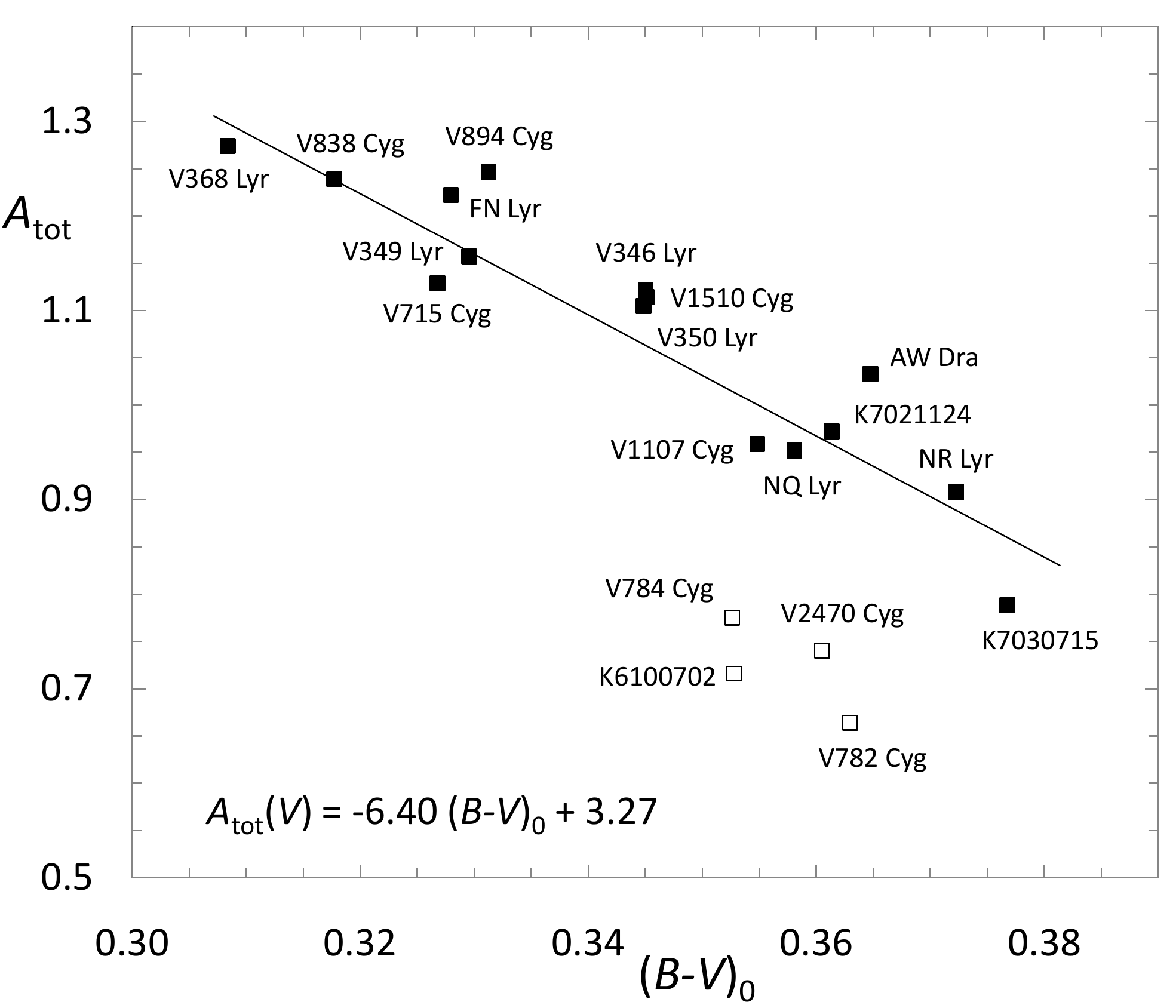} {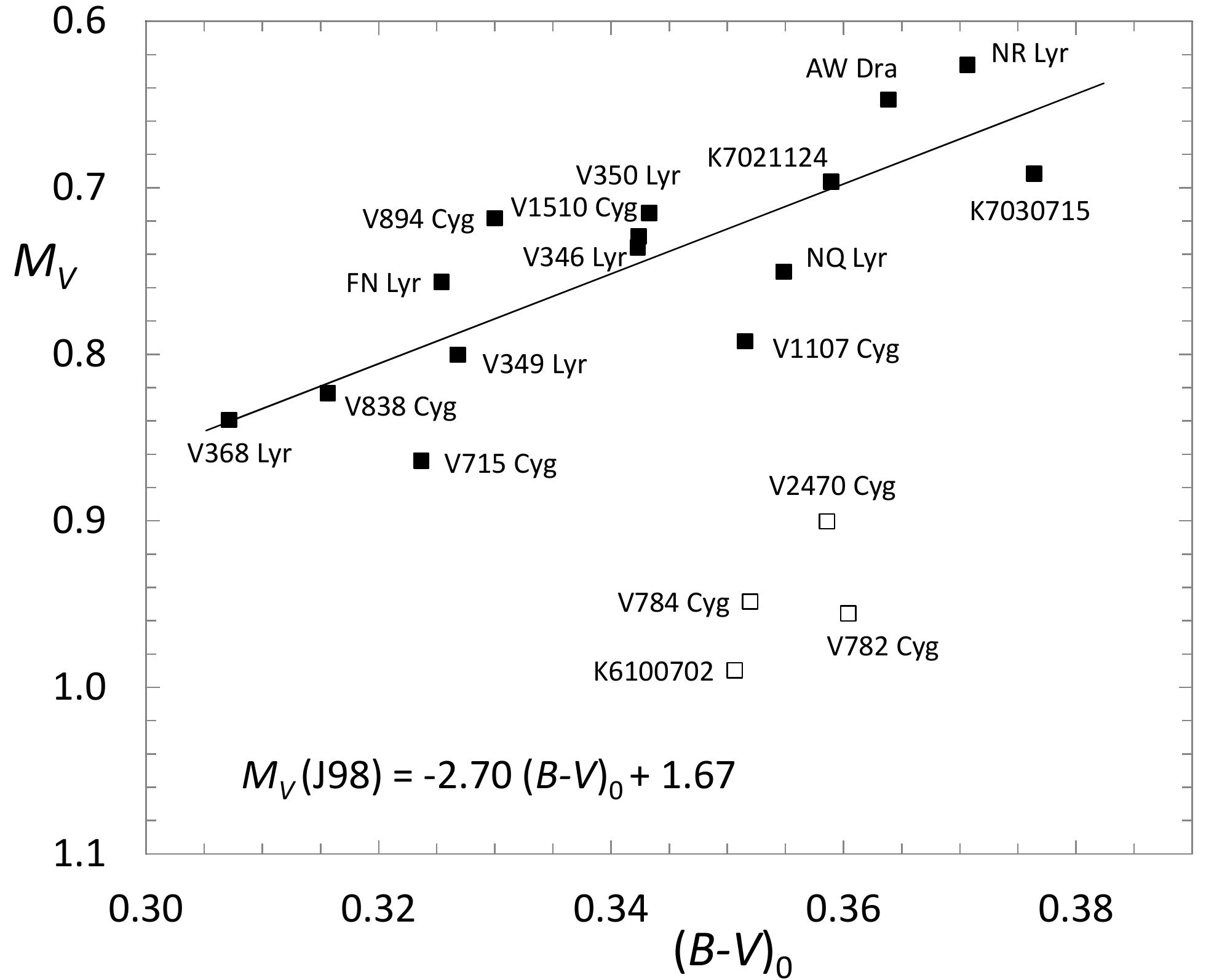} 
\vskip0pt
\caption{Four diagrams for the non-Blazhko RR~Lyr stars, all with mean dereddened colour  ($B-V$)$_0$
(average of the two values given in column~6 of Table~5)  along the abscissa:  
metallicity {\it vs.} colour (Top left), period {\it vs.} colour (Top right), 
total amplitude {\it vs.} colour (Bottom left), and absolute magnitude {\it vs.} colour (Bottom right).  
In each panel the line (and its equation) is from a least squares fit to the points for the 
15 metal-poor stars ({\it i.e.}, those with [Fe/H]$<-1.0$ dex).  }
\label{ColourRelationships}
\end{figure*}

{\bf Figure~7} summarizes the derived de-reddened colour information in the form of correlation diagrams.
In all four panels the four more metal rich stars stand apart from the more metal-poor stars.  For this reason
they have been plotted with open squares.   
The linear correlations show that the reddest metal-poor stars (those nearest the red edge
of the Instability Strip) are more metal poor, have longer periods and smaller total amplitudes, 
and are more luminous than the bluest metal-poor stars.
The diagrams also show that at a given colour the four metal-rich stars (plotted with open squares)  
have shorter periods, smaller amplitudes and are less luminous than the metal-poor stars.

Not represented in these diagrams but computed and given in the full paper are absolute magnitudes ($V$-passband),
surface gravities, mean effective temperatures, distances, and luminosities and masses.  Both 
pulsational and evolutionary  $L$ and $\cal M$ values were computed.  The $L$(puls) were calculated using 
equations 16 and 17 of J98, and in both cases the lower metallicity stars have the higher
$L$(puls).  The $\cal M$(puls) were calculated using equations 14 and 22 of J98. The average mass for
the four metal-rich stars is $\sim$0.50 $\cal M_{\sun}$ compared with the average mass for the
more metal-poor stars $\sim$0.60 $\cal M_{\sun}$.
Three different $L$(evol) were calculated, using equations 8, 10 and 12 of Sandage (2006). For the metal-poor
stars all three $L$(evol) are systematically larger than the $L$(puls) values.  For the
four metal-rich stars the agreement is better but there is a wide range of $L$(evol) owing to the
linear or non-linear [Fe/H] dependencies.  Regardless of which 
formula was used the most luminous stars have the lowest metallicities and there is 
internal consistence.  It is not clear whether the
$L$(puls) or $L$(evol) are correct. As was the case for the luminosities, the $\cal M$(evol)
are all larger than the corresponding $\cal M$(puls) values.

\begin{figure*}
\centering
\plottwo {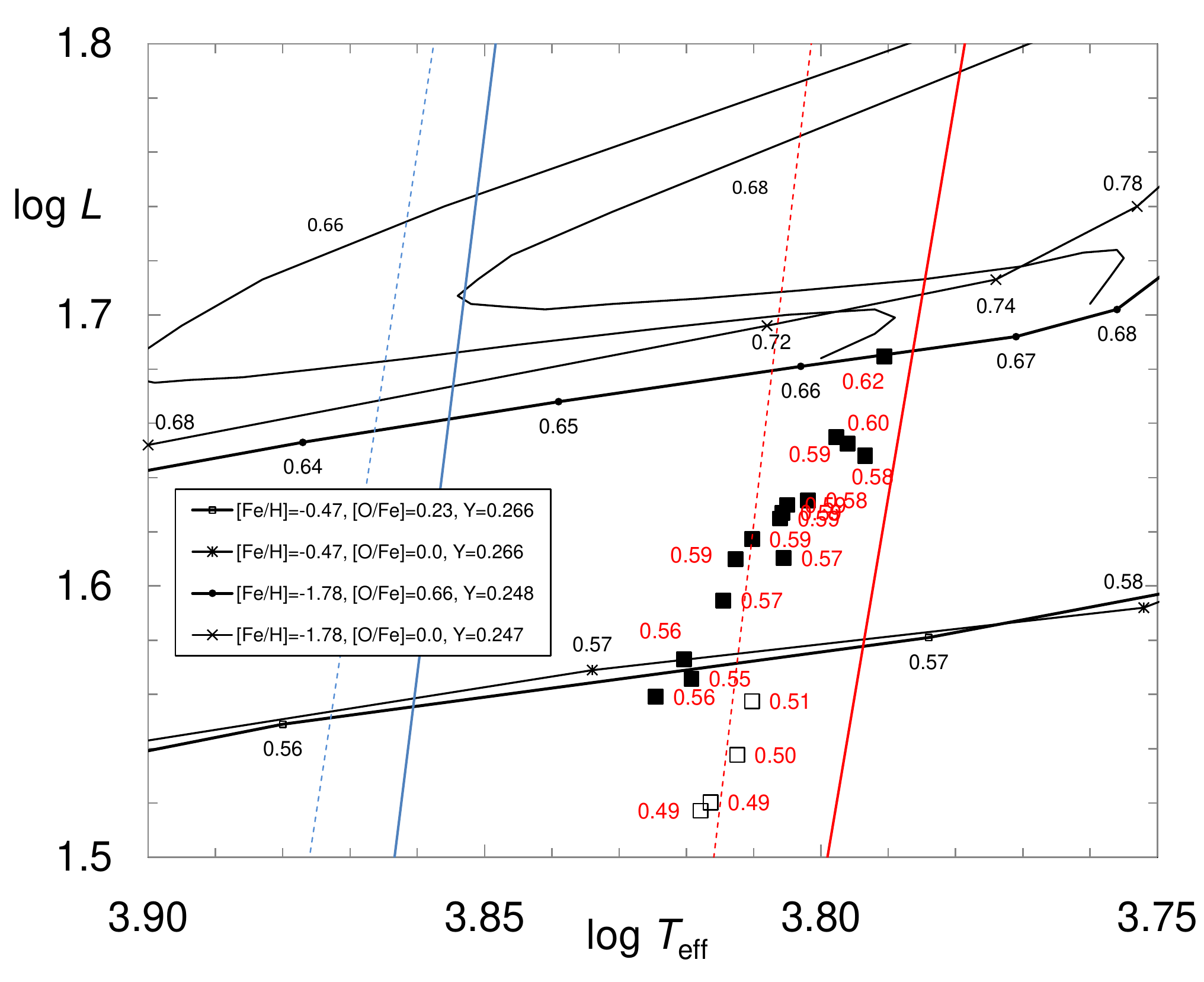} {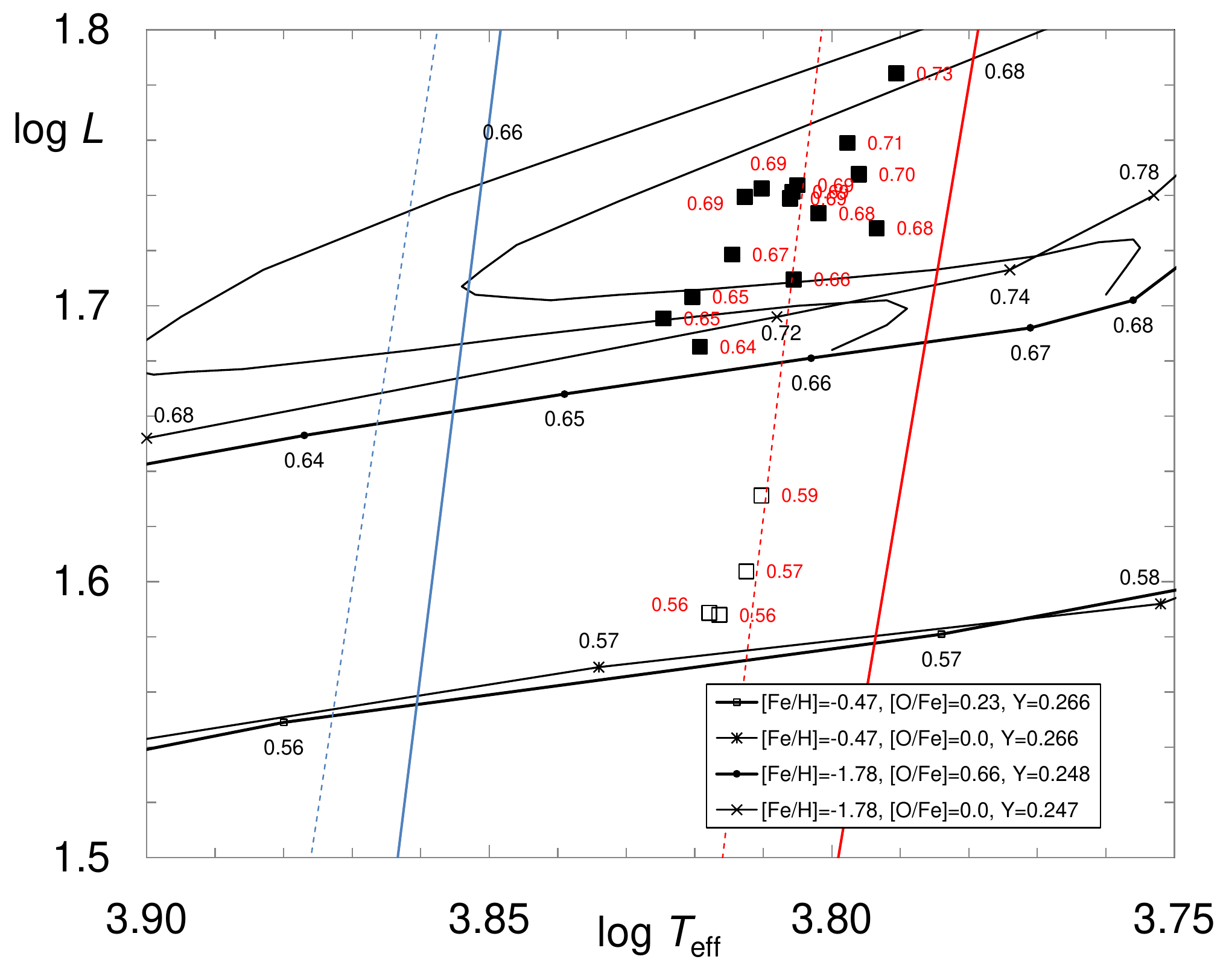} 
\plottwo {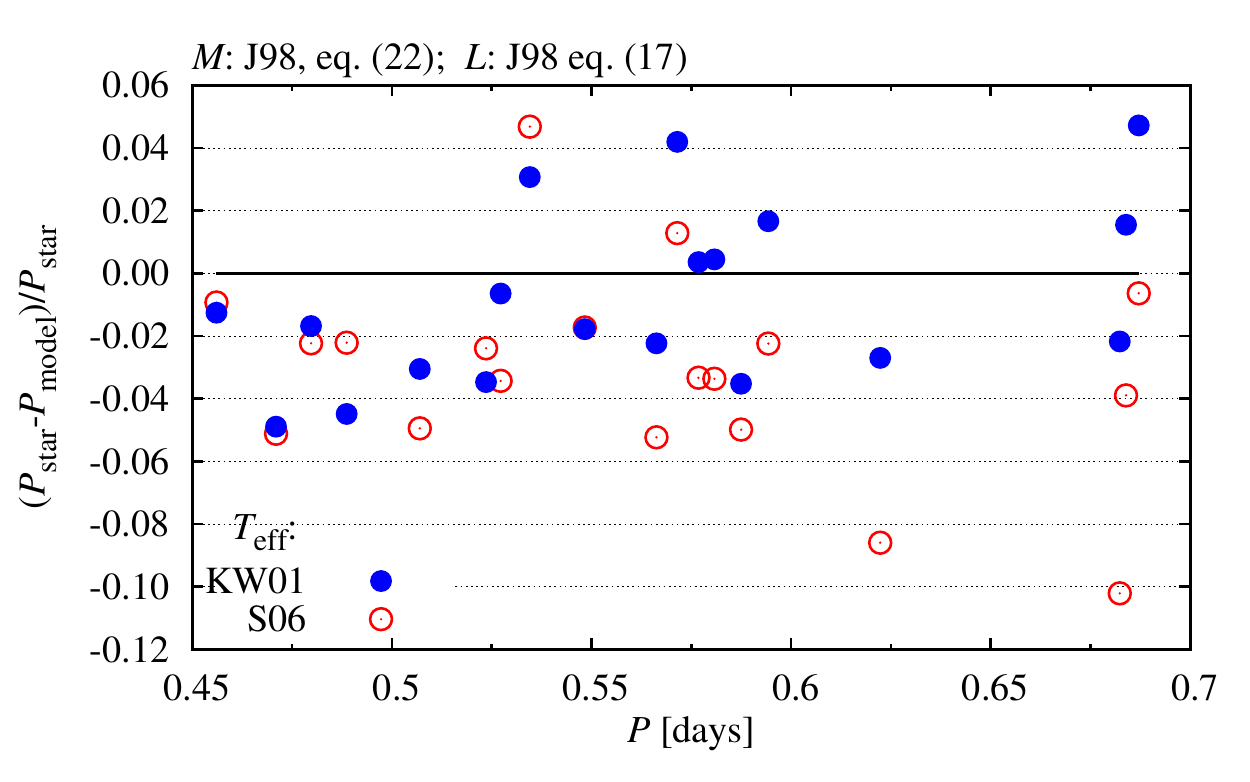} {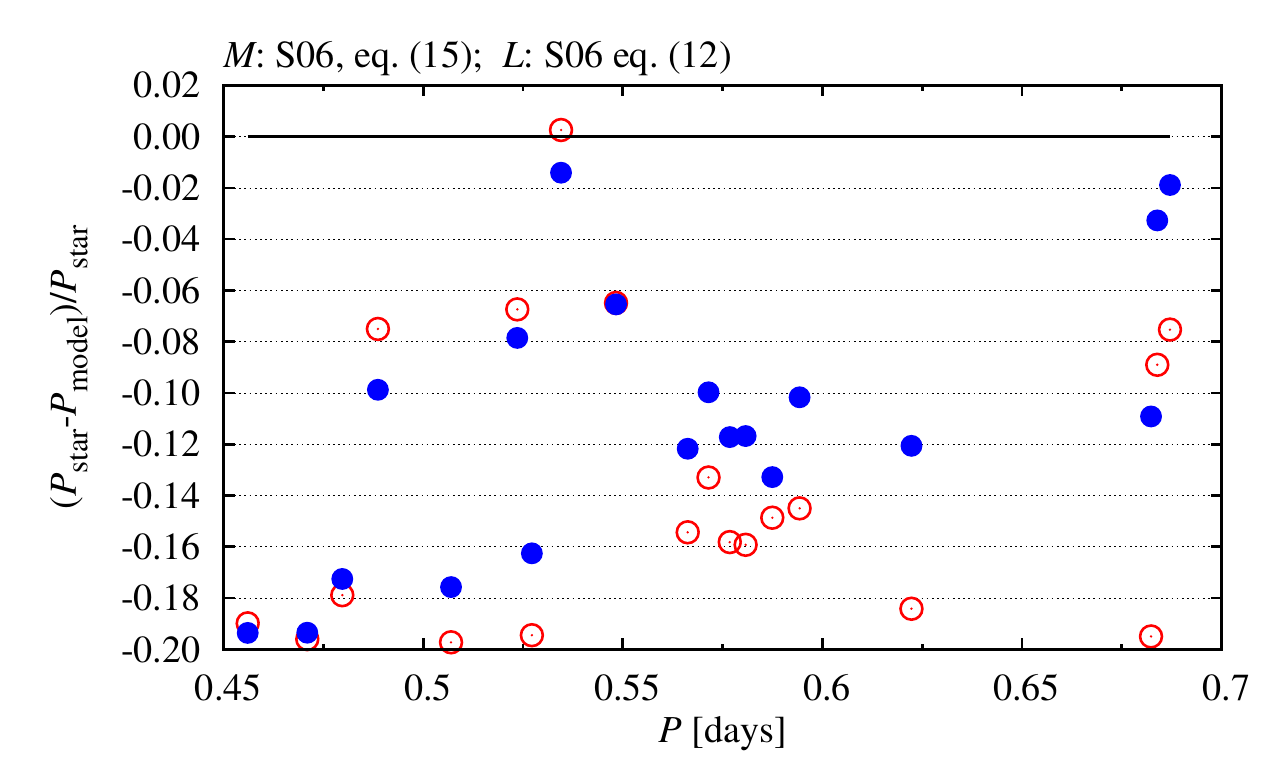}
\vskip0pt
\caption{
({\bf Top panels}) HR-diagrams showing the locations of the {\it Kepler} non-Blazhko RR~Lyr stars compared
with two sets of horizontal branch models (low- and high-metallicity) from Dorman (1992).  
In both graphs the four metal-rich RR~Lyr stars have been plotted with open squares and the metal-poor
stars with filled squares;  the black numbers next to the ZAHB symbols and the two evolutionary tracks are the 
assumed masses for the individual stellar evolution models.  
For the non-Blazhko RR~Lyr stars in the top left panel the $L({\rm puls})$ were calculated with 
eq.~17 of J98 and labelled (in red) with the ${\cal M}({\rm puls})$ calculated with
eq.~22 of J98;  in the top right panel the $L({\rm evol})$ were calculated with eq.~12 of S06 and labelled
(in red) with the ${\cal M}({\rm evol})$ calculated with eq.~15 of S06.
Both panels (left and right) show the blue and red edges for the instability strip
computed using the Warsaw pulsation code; the fundamental mode edges are plotted as solid lines, 
and first-overtone mode edges as dashed lines.  ({\bf Bottom panels}) Graphs comparing the observed pulsation periods and the 
pulsation periods derived with the Warsaw pulsation code. The assumed masses and luminosities
are the values given in the top panels.  For each star two points are plotted,
computed assuming the $T_{\rm eff}$ from eq.~11 of KW01 (blue dots) and from eq.~18 of S06 (red circles).
The best agreement is seen in the left panel, with the KW01 temperatures favoured over the S06 values.
In both panels the largest differences are seen for the longest period stars, with no metallicity
dependence. }
\label{HRdiags}
\end{figure*}

{\bf Figure 8} compares the results with the locations of model zero-age horizontal branches
and evolutionary tracks. 
The top two panels show HR diagrams with ${\rm log} \thinspace T_{\rm eff}$ as abscissa and ${\rm log} \thinspace L$ as ordinate, 
and two sets of ZAHB loci from Dorman (1992) computed for different masses along the ZAHB.  
The more luminous horizontal branch assumes [Fe/H]$=-1.78$ dex, and the less luminous branch [Fe/H]$=-0.47$ dex.
The numbers next to the symbols are the  masses for the individual models, which are seen to be higher for the
low-[Fe/H] tracks than for the high-[Fe/H] tracks.
For both assumed metallicities oxygen enhanced and non-enhanced ZAHBs have been plotted 
-- the effect of increasing the oxygen to iron ratio from [O/Fe]=0 to 0.66 for the low-metallicity ZAHBs 
is to lower the luminosity and reduce the mass at a given temperature.  
For the high-metallicity tracks an oxygen enhancement from [O/Fe]=0 to 0.23 has little
effect on the derived $L$ or ${\cal M}$.  Also plotted in the low-[Fe/H] case 
are the evolutionary paths away from the ZAHB for two masses, 0.66 and 0.68 ${\cal M}_{\odot}$.
In both panels the non-Blazhko RR~Lyr stars with low metallicities are represented by large black squares,
the four high-metallicity stars are plotted with open squares, 
and the $T_{\rm eff}$ are the average of the Kov\'acs \& Walker (2001, hereafter KW01) and Sandage (2006, hereafter S06) values.

In the top left panel of Fig.~8 the luminosities and masses (labelled in red) of the {\it Kepler} 
non-Blazhko RR~Lyr stars were calculated  
with eqs.~17 and 22 of J98 and thus are based on pulsation theory.  The $L$(puls) and ${\cal M}$(puls)
are seen to be systematically smaller than values derived from the ZAHB tracks 
(for the appropriate metal abundance).  The reddest non-Blazhko RR~Lyr stars lie
close to the fundamental mode red-edge and have the smallest amplitudes.  
This graph also shows blue and red edges of the instability strip for the 
fundamental mode (red and blue solid lines) and first-overtone mode (red and blue dashed lines).  The edges were calculated
with the Warsaw pulsation code assuming a mass of 0.65~${\cal M}_{\sun}$.  The {\it Kepler} non-Blazhko stars 
all lie in the fundamental mode region of the variability strip, and the smallest amplitude RR~Lyr stars 
(the four metal-rich stars, KIC~7030715 and NR~Lyr) have locations near the fundamental red edge (FRE) of the instability 
strip.  As expected, all the stars near the FRE have low $R_{31}$ values.

In the top right panel of Fig.~8  the luminosities were calculated with eq.~12 of S06, and the 
masses (labelled in red) with eq.~15 of S06; thus they are evolutionary $L$ and ${\cal M}$ values.  In this case there is 
very good agreement with the stellar evolution models, as one expects since they are based on stellar 
evolution models.  Enhancing the oxygen to iron ratio by the plotted amounts makes little difference. 

It is unclear which are correct, the $L$(puls) and ${\cal M}$(puls), or the $L$(evol) and ${\cal M}$(evol).   
The mass and luminosity discrepancies go in the same direction as seen for Cepheids.  
Pietrzynski {\it et al.} (2010) recently derived a dynamical mass ${\cal M}$(dynam)  
for a classical Cepheid in a well detached, double-lined eclipsing binary in the LMC.  The mass they derive 
is very accurate and favours ${\cal M}$(puls).   The reason for the discrepancies
may be the same, as suggested by Pietrzynski {\it et al}. -- not enough mass loss has been taken 
into account in the evolution models.    

Finally, the Nemec {\it et al.} (2011) paper concludes with a section on hydrodynamic models computed with the
Smolec and Moskalik `Warsaw convective pulsation program'.  Perhaps most relevant for the discussion above is the conclusion that in the $P$-$A$
diagram when [Fe/H] is varied from $-0.2$ to $-1.9$ dex there is very little motion at a given amplitude.  
Instead, motion to longer periods at a given amplitude occurs for stars of higher $L$ for
a given mass, or for stars of lower mass for a given luminosity.  This suggests that the main factors causing period
shifts in the $P$-$A$ diagram seem to be the mass and luminosity, the latter being caused
either by higher ZAHB mass or by post-ZAHB evolution.

\section{Conclusions}

The main conclusions reached by Nemec {\it et al.} (2011) concerning the {\it Kepler} non-Blazhko
RR~Lyrae stars are:  
(1) Fourier-based [Fe/H] values have been derived for
the stars (using four different correlations), and four of the 19 stars (V782~Cyg, V784~Cyg, KIC~6100702 and V2470~Cyg) 
appear to be much more metal-rich than the others -- recent CFHT spectra confirm that KIC~6100702 is metal-rich with [Fe/H]$=-0.18 \pm 0.06$ dex,
and AW~Dra is of intermediate metallicity with [Fe/H]=$-1.33 \pm 0.08)$ dex.
(2) The star KIC~7021124 is found to be doubly-periodic ($P_2/P_0 = 0.59305$) with characteristics
similar to V350~Lyr (see Benk\H o {\it et al.} 2010);
(3) All the stars show remarkably constant light curves ($\sigma < 1$ mmag) over the 
420-day interval of the Q0-Q5 observations;
(4) FN~Lyr and AW~Dra are found to have increasing periods; 
(5) the ASAS-North photometry of nine of the brightest stars shows that their colours are bluest when
brightest, as expected for RR~Lyrae stars;
(6) Fourier-based physical characteristics for all the stars have been derived;
(7) Distances range from 3 to 23 Kpc;
(8) Pulsational and evolutionary masses and luminosities are calculated for the stars,
the latter tending to be higher in both cases.   When the observed periods are compared with
periods computed with the Warsaw non-linear convective pulsation code better agreement
is achieved assuming {\it pulsational} $L$ and $\cal M$ values rather than the (higher) 
{\it evolutionary} $L$ and $\cal M$ values.  
(9) Finally, the Warsaw models show that varying [Fe/H] has a relatively small effect on
the $P$-$A$ relation and that the main factors causing the period shifts must be
luminosity and mass. 

\section{Acknowledgements}

Funding for the {\it Kepler} Discovery Mission is provided by NASA's Science Mission Directorate.  
We acknowledge the entire {\it Kepler} team for their outstanding efforts that have made these
results possible.  JMN thanks Dr. Andrew McWilliam for the invitation to present these results in Pasadena at such
a stimulating conference.

\end{document}